\documentclass[apjl]{emulateapj}

\usepackage{url}
\usepackage{color}
\usepackage{ulem}

\shorttitle{ALMA Observations of Cold Molecular Gas in NGC~1275}
\shortauthors{Nagai et al.}

\begin{document}


\title{The ALMA Discovery of the Rotating Disk and Fast Outflow of Cold Molecular Gas in NGC~1275}


\author{H. Nagai\altaffilmark{1,2}, K. Onishi\altaffilmark{3,14}, N. Kawakatu\altaffilmark{4}, Y. Fujita\altaffilmark{5}, M. Kino\altaffilmark{6,1}, Y. Fukazawa\altaffilmark{7}, J. Lim\altaffilmark{8,9}, W. Forman\altaffilmark{10}, J. Vrtilek\altaffilmark{10}, K. Nakanishi\altaffilmark{1,2}, H. Noda\altaffilmark{5}, K. Asada\altaffilmark{11}, K. Wajima\altaffilmark{12}, Y. Ohyama\altaffilmark{11}, L. David\altaffilmark{10}, \& K. Daikuhara\altaffilmark{13}}



\altaffiltext{1}{National Astronomical Observatory of Japan, Osawa 2-21-1, Mitaka, Tokyo 181-8588, Japan}\email{hiroshi.nagai@nao.ac.jp}
\altaffiltext{2}{The Graduate University for Advanced Studies (SOKENDAI), Osawa 2-21-1, Mitaka, Tokyo 181-8588, Japan}
\altaffiltext{3}{Research Center for Space and Cosmic Evolution, Ehime University, 2-5 Bunkyo-cho, Matsuyama, Ehime 790-8577, Japan}
\altaffiltext{4}{National Institute of Technology, Kure College, 2-2-11, Agaminami, Kure, Hiroshima, 737-8506, Japan}
\altaffiltext{5}{Department of Earth and Space Science, Graduate School of Science, Osaka University, 1-1 Machikaneyama-cho, Toyonaka-shi, Osaka 560-0043, Japan}
\altaffiltext{6}{Kogakuin University of Technology \& Engineering, Academic Support Center, 2665-1 Nakano, Hachioji, Tokyo 192-0015, Japan}
\altaffiltext{7}{Hiroshima University, 1-3-1 Kagamiyama, Higashi-Hiroshima 739-8526, Japan}
\altaffiltext{8}{Department of Physics, The University of Hong Kong, 0000-0002-6536-5575, Pokfulam Road, Hong Kong}
\altaffiltext{9}{Laboratory for Space Research, Faculty of Science, The University of Hong Kong, 0000-0002-6536-5575, Pokfulam Road, Hong Kong}
\altaffiltext{10}{Harvard-Smithsonian Center for Astrophysics, 60 Garden Street, Cambridge, MA 02138, USA}
\altaffiltext{11}{The Academia Sinica Institute of Astronomy and Astrophysics, AS/NTU. No.1, Sec. 4, Roosevelt Rd, Taipei 10617, Taiwan, R.O.C.}
\altaffiltext{12}{Korea Astronomy and Space Science Institute, 776 Daedeok-daero, Yuseong-gu, Daejeon 34055, Korea 0000-0001-6094-9291}
\altaffiltext{13}{Department of Physics, Toho University, 2-2-1 Miyama, Funabashi, Chiba 274-8510, Japan}
\altaffiltext{14}{Department of Space, Earth and Environment, Chalmers University of Technology, Onsala Observatory, SE-439 92 Onsala, Sweden}
%

\begin{abstract}
We present ALMA observations of the CO(2-1), HCN(3-2), and HCO$^{+}$(3-2) lines in the nearby radio galaxy / brightest cluster galaxy (BCG) of NGC~1275 with the spatial resolution of $\sim20$~pc.  In the previous observations, the CO(2-1) emission was detected as radial filaments lying in the east-west direction on kpc scale.  We resolved the inner filament and found that the filament cannot be represented by a simple infalling stream on sub-kpc scale.  The observed complex nature of the filament resembles the cold gas structure predicted by numerical simulations of cold chaotic accretion.  Within the central 100~pc, we detected a rotational disk of the molecular gas whose mass is $\sim10^{8}~M_{\sun}$.  This is the first evidence of the presence of massive cold gas disk on this spatial scale for BCGs.  A crude estimate suggests that the accretion rate of the cold gas can be higher than that of hot gas.    The disk rotation axis is approximately consistent with the radio-jet axis.  This probably suggests that the cold gas disk is physically connected to the innermost accretion disk which is responsible for jet launching.  We also detected absorption features in the HCN(3-2) and HCO$^{+}$(3-2) spectra against the radio continuum emission mostly radiated by $\sim1.2$-pc-size jet.  The absorption features are blue-shifted from the systemic velocity by $\sim$300-600~km~s$^{-1}$, suggesting the presence of outflowing gas from the active galactic nucleus (AGN).  We discuss the relation of the AGN feeding with cold accretion, the origin of blue-shifted absorption, and estimate of black hole mass using the molecular gas dynamics.
\end{abstract}		


\section{Introduction}\label{sect:intro}
The Perseus cluster is a nearby (z=0.017) cluster of galaxies and has been the subject of extensive research over years at all wavelengths.  The cluster harbors the radio galaxy / brightest cluster galaxy (BCG) NGC~1275, centered at the Perseus cluster.  Nuclear emission of NGC~1275 was originally classified as a Seyfert 2 \citep{Seyfert1943} but later has been also classified as a Seyfert 1.5/LINER \citep{Veron2006, Sosa-Brito2001} because of the presence of broad H$\alpha$ wing.  The total luminosity in 1990s was $4\times10^{44}$~erg~s$^{-1}$ \citep{Levinson1995}.  This luminosity is about 0.4\% of the Eddington luminosity for a black hole mass of $8\times10^{8} M_{\odot}$ \citep{Scharwachter2013}.  The radio luminosity of this source is $3\times10^{24}$~W~Hz$^{-1}$~sr$^{-1}$ at 178~MHz, which is classified as a Fanaroff-Riley I radio source \citep{Fanaroff1974}. 

This galaxy shows an intermittent jet activity in which the most recent activity started in $\sim2005$ \citep{Abdo2009, Nagai2010, Suzuki2012}.  This system is the brightest X-ray cluster of galaxies and one of the best examples of a prototypical ``cool core" cluster with a signature of the interaction between the jet and the intra-cluster medium (ICM) \citep[e.g.][]{Fabian2003}.  In contrast to other nearby BCGs such as M87 \citep{Tan2008, Perlman2007}, NGC~1275 is the reservoir of a large amount of cold gas ($M_{\rm gas}\simeq10^{10} M_{\sun}$) with filamentary structures on kpc scales \citep{Salome2006, Salome2008, Lim2008}, which is reminiscent of a cooling flow.  These properties make NGC~1275 the ideal target to study the feeding and feedback between the central super massive black hole (SMBH) and the surrounding environment.

Despite extensive efforts, the gas fueling mechanism from the ICM to the SMBH is not yet clear.  On kpc scales, molecular gas filaments are aligned in the east-west direction \citep{Lim2008} and those filaments seem to coincide with the H$\alpha$ nebulae \citep{Fabian2008}.  While the gas kinematics in the outer filaments is consistent with the infalling motion, the inner filament of the molecular gas \citep[$M_{\rm gas}\simeq10^{9} M_{\rm sun}$:][]{Lim2008} cannot be explained by either a simple inflow or outflow, and thus it is not clear if such a large amount of molecular gas is accreting to the center.  On the other hand, the circumnuclear disk (CND) is clearly resolved by the warm H$_{2}$ and ionized [Fe II] lines in the inner 50~pc, both in morphology and kinematics with the Gemini Telescope \citep{Scharwachter2013}.  It was suggested from the observed velocity dispersion that the H$_{2}$ emission traces the outer region of the disk which is likely to form a toroid while the [Fe II] line traces the inner region of the disk which is illuminated by the ionizing photons from the AGN.  The inner ionized part is possibly associated with the ``silhouette" disk, which is identified by the free-free absorption (FFA) of background synchrotron emission from the counter jet by VLBI observations \citep[e.g][]{Walker2000, Fujita2017}.  \cite{Fujita2017} measured the opacity of free-free absorption ($\tau_{\rm ff}$) toward the counter jet component in the central $\sim1$~pc region and found $\tau_{\rm ff}\propto\nu^{-0.6}$, which is different from that for a uniform density of $\tau_{\rm ff}\propto\nu^{-2}$.  They argued that the absorbing medium is highly inhomogeneous and that it consists of regions of $\tau_{\rm ff}\ll 1$ and $\tau_{\rm ff}\gg 1$.  

Numerical simulations of giant elliptical galaxies suggest that the mass accretion is dominated by chaotic cold accretion (CCA) within the inner kpc \citep{Barai2012, Gaspari2013}.  These simulations also predict that the CCA leads to a deflection of jets and strong variation in the AGN luminosity.  Such a jet deflection and luminosity change are indeed observed in NGC~1275/3C~84 in radio, X-ray, and $\gamma$-ray bands \citep{Nagai2010, Dutson2014, Fabian2015}, though the connection between different frequency bands is not well understood \citep{Nagai2012, Nagai2016, Hodgson2018}.  Thus, Gaspari's simulation seems to represent the observed AGN properties of NGC~1275.  Contrary to other elliptical galaxies in a cool-core cluster system such as M87 \citep{Forman2007}, NGC~1275 indeed possesses a large amount of cold gas in the center \citep[M$_{\rm gas}\sim10^{9}$M$_{\rm sun}$:][]{Lazareff1989, Mirabel1989, Lim2008, Salome2008}.  \cite{Salome2011} discussed a possibility of the presence of CO molecular disk within 2~kpc.  However, resolving the inner ($<1$~kpc) region has so far been difficult because of the lack of resolution at mm wavelengths.  High resolution observations of the cold molecular gas with ALMA are key to uncovering the missing link between the outer kpc and the inner $\lesssim50$~pc, which is crucial to understand the mass accretion from the ICM to SMBH. 

In broader perspectives, it is also intriguing to study the relation of the observed CND with the AGN torus in terms of the unified scheme \citep{Antonucci1993}.  Although many studies have been done for prototypical Seyfert galaxies with ALMA in this regard \citep[e.g.][]{Imanishi2018}, the properties of the CND and its relation to the AGN torus in BCG/radio galaxies is less clear.  From theoretical studies, it is predicted that the thickness of the CND can vary with the accretion rate \citep[e.g.][]{Kawakatu2008}.  It should be noted that the Eddington ratio of NGC~1275 ($\sim10^{-3}$) is close to the boundary between Seyfert AGN and low luminosity AGN, which are thought to have a standard disk and RIAF-type accretion flows \citep{Narayan1994, Yuan2014}, respectively.  Thus, the disk thickness of NGC~1275 may be different from that in typical Seyfert AGNs.

In the cosmological context, BCGs provide a unique opportunity to test the hierarchical galaxy formation scenario along with the cluster formation.  BCGs are the most luminous and massive galaxies in the Universe at the present epoch and typically located near the center of their parent clusters.  They are basically interpreted to form through the merger of several massive galaxies in the cluster's history \citep{Dubinski1998, DeLucia2006, Tonini2012}.  Galaxy mergers are thought to grow the SMBH and drive the scaling relation between SMBHs and their host galaxies \citep[e.g.][]{Kormendy2013, McConnell2013}.  It is therefore natural to expect that the most massive SMBHs reside in BCGs.  ALMA can probe molecular gas dynamics within the sphere of gravitational influence (SoI) for the nearby BCGs including NGC~1275, which allows us to measure the SMBH mass. 

In this paper, we present the ALMA observations of NGC~1275 using the CO(2-1), HCN(3-2), and HCO$^{+}$(3-2) lines to study the cold gas morphology and kinematics from sub-kpc to $\sim30$~pc scale.  Throughout this paper, we use $H_{0}$=70.5, $\Omega_{\rm M}=0.27$, and $\Omega_{\Lambda}=0.73$.  At the 3C 84 distance, $0.1\arcsec$ corresponds to 34.4~pc.

\section{OBSERVATIONS and DATA ANALYSIS}\label{sect:Obs}
We proposed ALMA observations to study the morphological and kinematical properties of the cold molecular gas in the circumnuclear region ($\sim100$~pc) of NGC~1275 in ALMA Cycle 5 which was approved as the project code of 2017.1.01257.S.  The proposal consists of two Group Observing Unit Sets (OUSs), the one for the observation of the CO(2-1) transition and the other for the observation of the HCN(3-2) and HCO$^{+}$(3-2) transitions.  Each Group OUS splits into two Member OUSs where the one is for the observation with an extended configuration (C43-7) and the other is for the observation with a compact configuration (C43-4).  This paper presents the results from the observations with the extended configuration.

The observations were done in ALMA Band 6 ($\lambda=$1.3~mm) with 47 antennas on 2017 November 27 and 2017 November 28 for the CO(2-1) and HCN(3-2)/HCO$^{+}$(3-2), respectively.  The maximum and minimum baseline lengths are 2.5~km and 92.1~m. respectively.  The observations consist of 52 scans for the CO(2-1) and 46 scans for the HCN(3-2)/HCO$^{+}$(3-2) with a 54.4-sec integration for each scan bracketed by the scans for the complex gain calibrator observations.  Integration time per an interferometric visibility was set to 2 sec.  Table \ref{tab:Obs} summarizes the observations.  Both X and Y linear polarizations were received and parallel-hand correlations (XX and YY) were obtained by the ALMA 64-input correlator.  For the CO(2-1) observation, two spectral windows (spws) were set in both the lower sideband (LSB) and the upper sideband (USB), with 1920 channels per spw (0.977~MHz channels resolution or $\sim$1.3~km/s velocity resolution) and 64 channels per spw (15.625~MHz channels resolution or 19.5 km/s velocity resolution) for the LSB and USB, respectively.  One of the spws in the LSB was centered at the frequency slightly shifted from the expected CO(2-1) frequency so that another spw in the LSB can partially cover the CN(2-1) line \footnote{The CN(2-1) is not detected from our observations.  The spectral dynamic range should be similar to that of the CO(2-1) spw.  Thus, we can give an upper limit of $\sim6.8$~mJy (see Table 3).}.  For the HCN(3-2) and HCO$^{+}$(3-2) observations, two spectral windows (spws) were set in both the LSB and the USB, with 64 channels per spw (15.625~MHz channels resolution or 19 km/s velocity resolution) and 1920 channels per spw (0.977~MHz channels resolution or $\sim$1.1~km/s velocity resolution) for the LSB and USB, respectively.  Two spws in the USB were centered at the HCN(3-2) and HCO$^{+}$(3-2) lines, respectively.

The data were processed at EA-ARC with the CASA 5.1.1-5 and ALMA Pipeline version 40896.  Amplitude calibration was performed using measurements of the system
Temperature (T$_{\rm sys}$) on a channel-by-channel basis.  Rapid phase variations on timescales of less than the gain calibration cycle were corrected using the water vapor radiometer. Unreliable data such as amplitude phase jumps and low antenna gains were automatically identified by the ALMA Pipeline and flagged.  The bandpass calibration was done both in phase and amplitude. Temporal variation of the gain amplitude and phase was calibrated using the averaged XX and YY correlations of a gain calibrator.  Flux scaling was derived on the bandpass calibrator using the flux information provided by the Joint ALMA Observatory (JAO).  The referenced flux was derived with the interpolation (power-law spectral index of $-0.63$) from measurements in Bands 3 and 7 on 2017 November 27. 

After applying all calibrations, it was found that the pipeline calibration was not sufficient to correct the phase delay, which was seen in both gain calibrator and NGC~1275. The residual delay was antenna-dependent and time variable from scan to scan, which showed variations within at most $\pm20$ psec.  This amount of delay can cause the phase slope of a few degrees across one spw.  Thanks to bright continuum emission of NGC~1275 ($\sim$7~Jy), we calibrated the residual delay using the CASA task \texttt{gaincal} with \texttt{gaintype=K}.  Since NGC~1275 can be observed at low elevation from the ALMA site, the visibility phase showed very rapid time variation, which cannot be fully corrected with usual gain calibration nor WVR correction.  Thus, we performed self-calibration using the continuum emission of NGC~1275 in both phase and amplitude.  The phase and amplitude self-calibration were done per integration time and scan, respectively.

The continuum emission was identified at line-free channels with the fitting order of one using the CASA task \texttt{uvcontsub}.  Images were created with the velocity resolution of 20~km/s with the ``nearest" interpolation in frequency.  Deconvolution was done with the CLEAN algorithm using the CASA task \texttt{tclean} non-interactively.  In the following sections, we show images with both naturally weighting and Briggs weighting \citep{Briggs1995} of robust parameter 0.5.  The beam shapes for all images are summarized in Table \ref{tab:beam}.  Those for archival data (see \S\ref{sect:HCN-HCO}) are also summarized in the same Table.

We note that some residual emissions at the image center are seen at the line-free channels even after the continuum subtraction.  Those emissions are not constant over the channels but appears almost randomly from channel to channel.  This is most likely due to the limitation of spectral dynamic range caused by the bandpass calibration uncertainty.  Assuming that the bandpass accuracy depends on the signal-to-noise ratio of bandpass calibrator and the residual after the bandpass calibration is random over antennas and frequency, we can estimate the spectral dynamic range $D_{\rm spec}$ as $D_{\rm spec}=I_{\rm BP}/\sigma_{\rm BP}$ where $I_{\rm BP}$ is the flux density of bandpass calibrator and $\sigma_{\rm BP}$ is the sensitivity for the bandpass calibrator scan.  We estimate $\sigma_{\rm BP}$ using the measured image rms on the bandpass calibrator image.  
The image of the target source should also exhibit the error given by $\sigma_{\rm T}=I_{\rm T}/D_{\rm spec}$ after the continuum subtraction, where $I_{\rm T}$ is the continuum flux density of target source.  Table \ref{tab:SpecDR} summarizes the value of these parameters.  The estimated spectral dynamic range is nearly consistent with the demonstrated spectral dynamic range of $1000$ given in the ALMA Proposer's Guide and Technical Handbook \citep{Andreani2018, Warmels2018}.  Using the measured continuum emission for I$_{\rm T}$, we obtain an uncertainty ($\sigma_{\rm T}$) of 6.8~mJy, 5.5~mJy, and 6.3~mJy 
for the CO(2-1), HCN(3-2), and HCO$^{+}$(3-2) lines, respectively.  These uncertainties dominate the off-position image rms of $\sim1$~mJy (see section \ref{sect:results}).  In the following sections, we will show the observed spectra integrated over the aperture of $0.8\arcsec$-diameter circle and $0.3\arcsec$-diameter circle.  We propagate the above uncertainties into our analysis for the spectra.  We obtain $\sigma_{\rm T}$ of 4.3~mJy for both the HCN(3-2) and HCO$^{+}$(3-2) for the archival data (see \S\ref{sect:HCN-HCO}).  The reason for a better $\sigma_{\rm T}$ in the archival data is that the data was taken in relatively good weather condition in terms of the T$_{\rm sys}$ and WVR, and the bandpass calibrator is brighter so that it provides a better signal-to-noise ratio for the bandpass calibration.  
For an additional validation, we measured the image rms within the same aperture used for Figure \ref{fig:COspectrum} after the continuum subtraction for the line-free spws.  The measured rms is about 0.9~mJy, which is smaller than $\sigma_{\rm T}$.  Thus, we confirmed that the baseline oscillation in line-free spws does not exceed the estimated uncertainties.

\begin{table*}
\caption{Observation Summary}\label{tab:Obs}
  \begin{center}
    \begin{tabular}{ccccccc} \hline\hline
Target Lines & ToS [min]\tablenotemark{a} & NoA\tablenotemark{b} & Bandpass\tablenotemark{c} & Gain\tablenotemark{d} & T$_{\rm sys}$\tablenotemark{e} & MRS\tablenotemark{f} \\ \hline
CO(2-1) & 45 & 47 & J0237+2848 & J0313+4120 & 100~K & $1.8\arcsec$ (610~pc) \\
HCN(3-2), HCO$^{+}$(3-2) & 39 & 47 &  J0237+2848 & J0313+4120 & 120~K & $1.5\arcsec$ (530~pc) \\ \hline
\end{tabular}
\end{center}
\tablenotetext{1}{Total integration time on NGC~1275}
\tablenotetext{2}{Number of antennas participated to the observation}
\tablenotetext{3}{Bandpass calibrator. Note that bandpass calibrator is also used as the flux calibrator to derive the amplitude scaling.}
\tablenotetext{4}{Complex gain calibrator}
\tablenotetext{5}{Typical system temperature}
\tablenotetext{6}{Maximum recoverable scale given by $0.6\lambda/D_{\rm min}$ \citep[see][]{Warmels2018}, where $D_{\rm min}$ is the minimum baseline length.}
\end{table*}

\begin{table*}
\caption{Beam Shape}\label{tab:beam}
  \begin{center}
    \begin{tabular}{ccccc} \hline\hline
Target Lines & Natural\tablenotemark{a} & Briggs\tablenotemark{b}  \\ \hline
CO(2-1) & $0.144\arcsec\times0.077\arcsec$ at $4.0\degr$ & $0.110\arcsec\times0.048\arcsec$ at $0\degr$ \\
HCN(3-2) & $0.129\arcsec\times0.076\arcsec$ at $12.1\degr$ & $0.098\arcsec\times0.045\arcsec$ at $9.4\degr$ \\
HCO$^{+}$(3-2) & $0.129\arcsec\times0.076\arcsec$ at $12.0\degr$ & $0.099\arcsec\times0.045\arcsec$ at $9.3\degr$ \\  
HCN(3-2) (archival data) & -\tablenotemark{c} & $0.901\arcsec\times0.363\arcsec$ at $-0.5\degr$ \\
HCO$^{+}$(3-2) (archival data) & -\tablenotemark{c} & $0.919\arcsec\times0.371\arcsec$ at $0.5\degr$ \\\hline
\end{tabular}
\end{center}
\tablenotetext{1}{The beam shape for the images with natural weighting.}
\tablenotetext{2}{The beam shape for the images with Briggs weighting (robust$=0.5$).}
\tablenotetext{3}{We use only the images with Briggs weighting for the spectral measurement of archival data.}
\end{table*}

\begin{table*}
\caption{Spectral Dynamic Range}\label{tab:SpecDR}
  \begin{center}
    \begin{tabular}{ccccccc} \hline\hline
Target Lines & $\nu_{\rm rest}$ [GHz] \tablenotemark{a} & I$_{\rm BP}$ [Jy]\tablenotemark{b} & $\sigma_{\rm BP}$ [mJy]\tablenotemark{c} & D$_{\rm spec}$\tablenotemark{d} & I$_{\rm T}$ [Jy]\tablenotemark{e} & $\sigma_{\rm T}$ [mJy]\tablenotemark{f}  \\ \hline
CO(2-1) & 230.538 & 1.082	& 1.03	& 1050 &	7.15 & 6.81 \\
HCN(3-2) & 265.886 & 0.989 & 0.76 & 1301 & 7.14 & 5.49 \\
HCO$^{+}$(3-2) & 267.558 & 0.985 & 0.87 & 1132 & 7.13 & 6.30 \\ 
HCN(3-2) (archival data) & 265.886 & 1.595 & 0.75 & 2141 & 9.26 & 4.33 \\
HCO$^{+}$(3-2) (archival data) & 267.558 & 1.590 & 0.75 & 2134 & 9.26 & 4.34 \\\hline
\end{tabular}
\end{center}
\tablenotetext{1}{Rest frequency of the line}
\tablenotetext{2}{Flux density of bandpass calibrator (J0237+2848)}
\tablenotetext{3}{1-$\sigma$ sensitivity for bandpass calibrator (J0237+2848) scan}
\tablenotetext{4}{Spectral dynamic range}
\tablenotetext{5}{Flux density of target source (NGC~1275)}
\tablenotetext{6}{Expected error for target source (NGC~1275)}
\end{table*}

\section{RESULTS}\label{sect:results}
\subsection{The CO(2-1) Line}
\begin{figure*}
  \begin{center}
   \includegraphics[width=170mm]{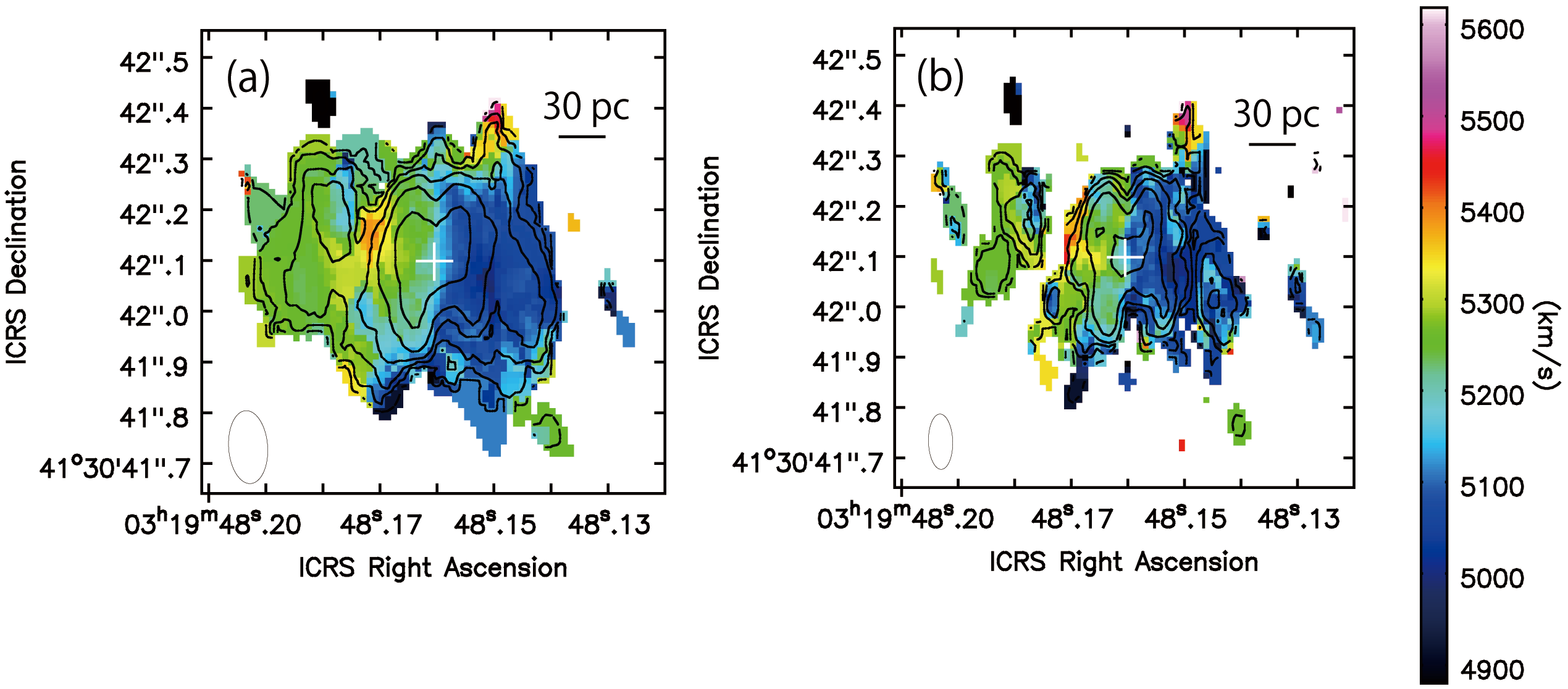}
 \end{center}  
\caption{(a) Contours of the naturally-weighted moment 0 map overlaid on the color of moment 1 map of CO(2-1).  The cross symbol indicates the AGN position, which was identified as the peak position of the continuum emission.  (b) The same as (a), but the image is created using the Briggs weighting with the robust parameter of 0.5.  The images are plotted through out the region where the flux density is greater than 5$\sigma$ ($1\sigma=0.87$~mJy for the naturally weighted image and $1\sigma=0.79$~mJy for the Briggs weighted image).  The contours are plotted at the level of $20$~km~s$^{-1}\times3\sigma\times (-1, 1, 2, 4, \cdots, 512)$.  The image rms is measured within a circle with radius of $1.8\arcsec$ (150 pixels) at the image center at line-free channels (averaged over 30-channels around $6000$~km~s$^{-1}$.).   The beam size is $0.144\arcsec \times 0.077\arcsec$ at the position angle of $4.0^{\circ}$ and $0.110\arcsec \times 0.048\arcsec$ at the position angle of $0^{\circ}$ for natural weighting and Briggs weighting, respectively.}
  \label{fig:CO}
\end{figure*}

\begin{figure}
  \begin{center}
   \includegraphics[width=85mm]{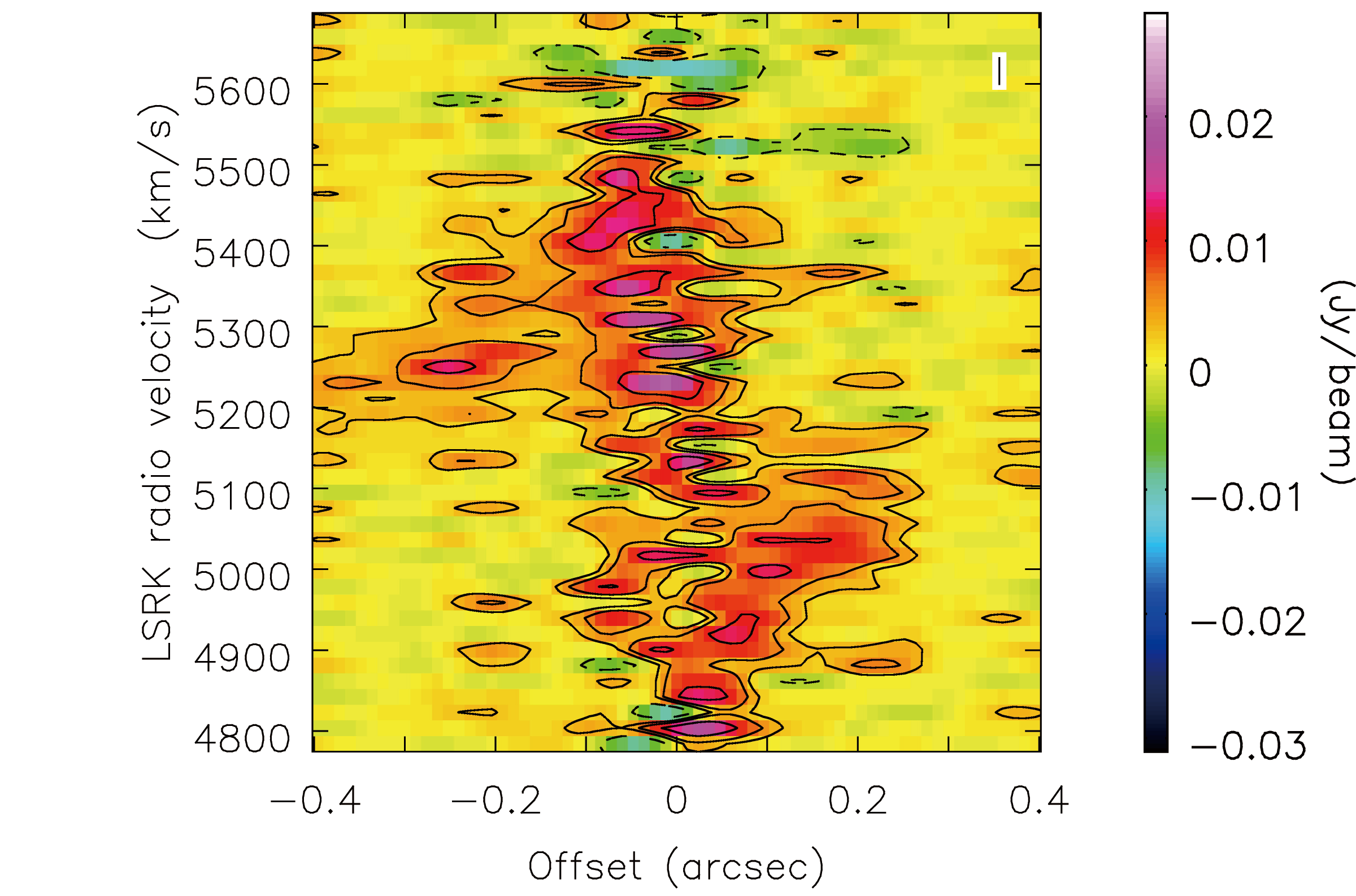}
 \end{center}  
\caption{Position-Velocity diagram of the CO(2-1) along the axis going through the image center of Fig. \ref{fig:CO} in a position angle of $70^{\circ}$.   }
  \label{fig:CO_PVD}
\end{figure}

\begin{figure}
  \begin{center}
   \includegraphics[width=85mm]{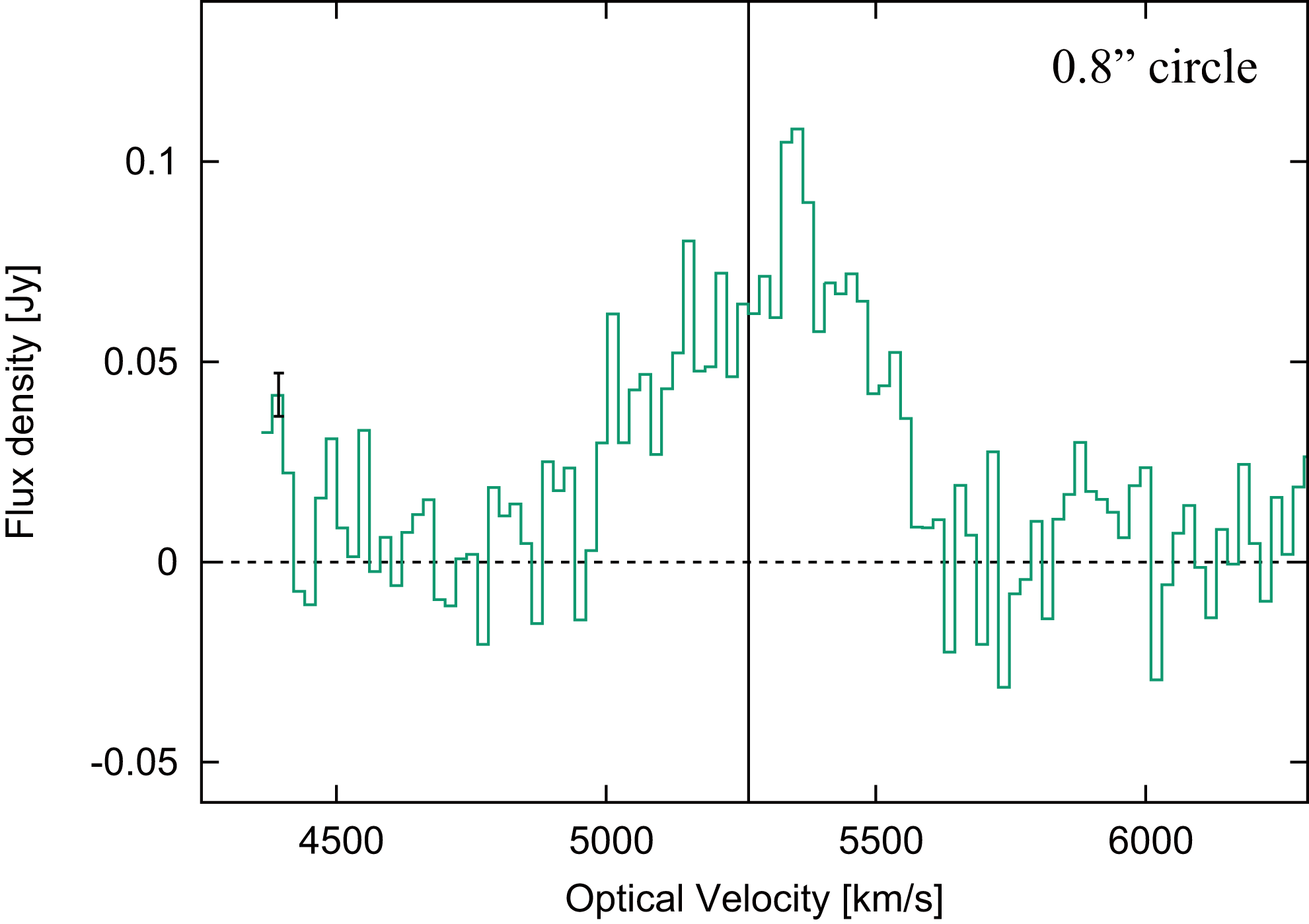}
 \end{center}  
\caption{Integrated spectrum of the CO(2-1) over the region of $0.8\arcsec$-diameter circle at the galaxy center.  An error bar corresponding to $\pm1\sigma$ uncertainty (6.8~mJy) is plotted in the panel.  The error estimation is explained in the text (see \S\ref{sect:Obs}).  The vertical line indicates the systemic velocity of NGC~1275.}
  \label{fig:COspectrum}
\end{figure}

\begin{figure*}
\begin{center}
\includegraphics[width=18.5cm]{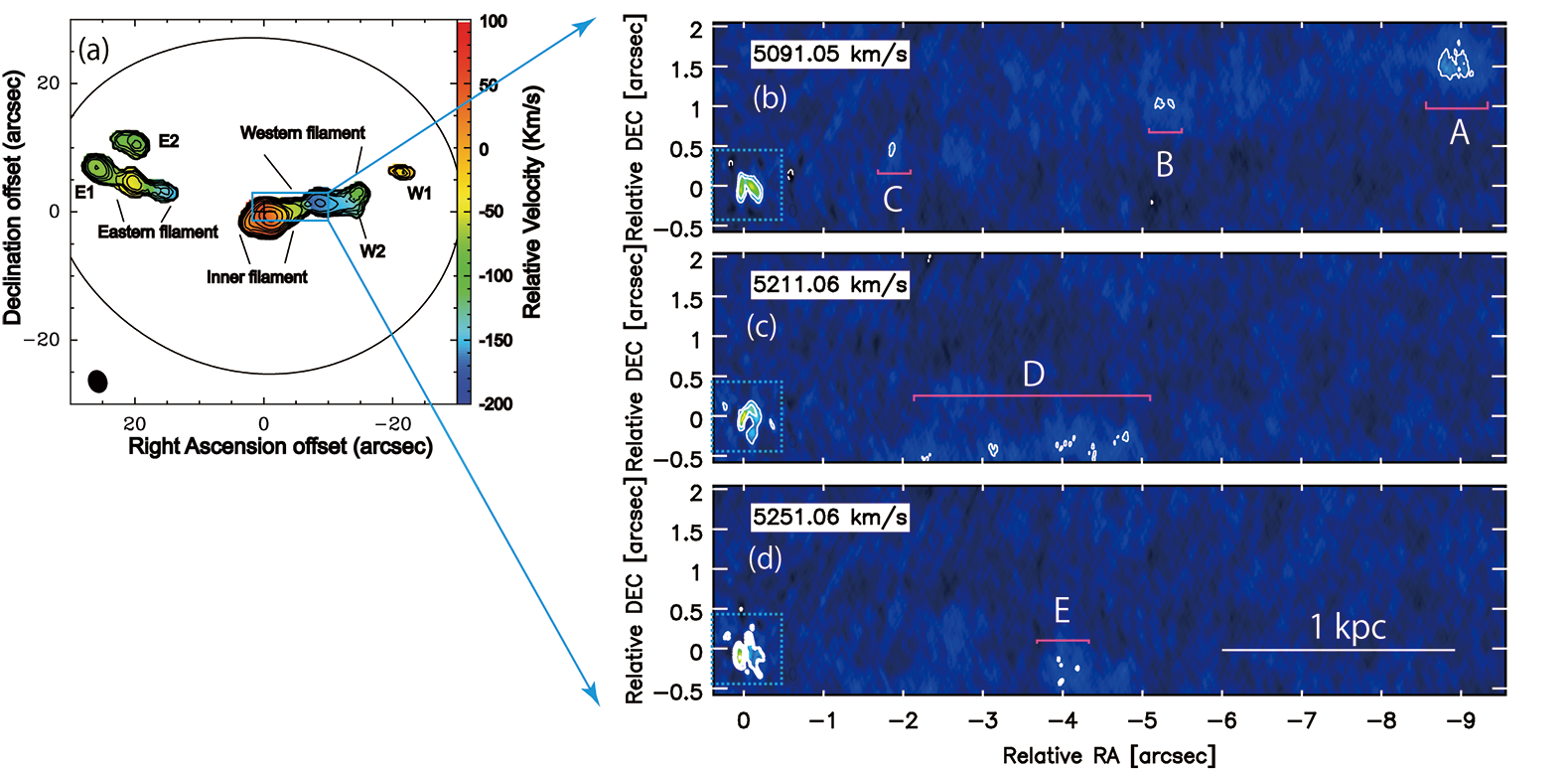}
\caption{(a) Contours of the moment 0 map overlaid on the moment 1 map in color by previous SMA observations \citep{Lim2008} (Reproduced by permission of the AAS.).  (b)-(d) Large-scale channel map of the CO(2-1) intensity by our ALMA observations at $v=5091$ km~s$^{-1}$ ($v-v_{sys}=-173$~km~s$^{-1}$), 5211 km~s$^{-1}$ ($v-v_{sys}=-53$~km~s$^{-1}$), and 5251 km~s$^{-1}$ ($v-v_{sys}=-13$~km~s$^{-1}$).  The images were created with natural weighting.  The contours are plotted at $5\sigma\times (-1, 1, 2, 4, \cdots, 32)$, where $1\sigma=0.87$~mJy.  The locations where the large-scale emissions are detected are highlighted by the magenta lines.  The rectangle shown in cyan lines in (a) indicates the region where the ALMA images are shown in (b)-(d).  The rectangle shown in broken cyan lines in (b)-(d) corresponds to the region shown in Figure 1.}
\label{fig:CO_Chan}
\end{center}
\end{figure*}

Figure \ref{fig:CO}(a) shows the contour of naturally-weighted velocity-integrated map (moment 0 map) overlaid on the velocity-weighted intensity map (moment 1 map) of the CO(2-1) line.  The emissions are mostly detected within a radius of 100~pc.  The structure seems to show a disk-like morphology.  We can see a clear velocity gradient with about $\pm250$~km~s$^{-1}$ in a position angle of $\sim70^{\circ}$ (Figure \ref{fig:CO_PVD}).  The intensity-weighted velocity at the AGN position is about 5200~km~s$^{-1}$, which is consistent with the optical systemic velocity.   The velocity gradient likely originates in the rotational motion of the gas disk, which is also evident in the HCN(3-2) and HCO$^{+}$(3-2) emissions (see next subsection).  The integrated spectrum over the central 0.8$\arcsec$ is shown in Figure \ref{fig:COspectrum}.  The emission is centered nearly at the systemic velocity, ranging from $\sim5000$~km~s$^{-1}$ ($v-v_{\rm sys}\simeq-264$~km~s$^{-1}$) to $\sim5600$~ km~s$^{-1}$ ($v-v_{\rm sys}\simeq336$~km~s$^{-1}$).  Previous Gemini observations of the warm H$_{2}$ gas \citep{Scharwachter2013} revealed a similar rotating disk structure within a radius of $\sim0.15\arcsec$ ($\sim50$~pc).  The line width of the warm H$_{2}$ gas shows $\sim350$~km~s$^{-1}$ of FWHM.  Both morphological and kinematical structures of CO(2-1) as well as its spectral shape are quite consistent with those for the warm H$_{2}$ gas, but our observed disk seems to have clumpy or filamentary substructures (see the higher angular resolution image created using the Briggs weighting with the robust parameter of 0.5, Figure \ref{fig:CO}(b)).  The CO counterpart of the H$_{2}$ gas stream, the south-west from the center \citep{Scharwachter2013}, is not detected in our images.  We note that there should be missing flux of extended structure in our data because of the lack of short interferometric baselines, so the whole structure of the disk can be larger.  The peak flux ($\sim0.1$~Jy) of the integrated spectrum (Figure \ref{fig:COspectrum}) is about a half of that of the ``inner filament" (see Fig. \ref{fig:CO_Chan}(a)) presented in \cite{Lim2008}.    The peak of the spectrum is at 5340~km~s$^{-1}$, which is about 100~km~s$^{-1}$ redshifted from the systemic velocity, while the spectral shape of the inner filament in \cite{Lim2008} is rather symmetric and centered at the systemic velocity.  This indicates that the blue-shifted emission is resolved-out even more strongly in our observations.

Some extended features are marginally detected in the western region in the naturally-weighted image (Figure \ref{fig:CO_Chan}).  Three components (denoted A, B, and C in Figure \ref{fig:CO_Chan}(b)) are mainly detected at $\sim8\arcsec$, $\sim5\arcsec$, and $\sim2\arcsec$ from the center in a position angle of $-80^{\circ}$ at $v\sim5091$~km~s$^{-1}$ ($v-v_{\rm sys}\simeq-173$~km~s$^{-1}$).  Those components probably represent dense, compact regions within the filaments reported by \cite{Lim2008}.  One of the outer components (A) was identified as a prominent knot-like feature with relative velocity of $\sim-150$~km~s$^{-1}$ in the western filament in Lim et al. (2008).  The inner components (B and C) seems to constitute the western elongation of the inner filament.  Another features (denoted D in Figure \ref{fig:CO_Chan}(c)) is marginally detected at $\sim4$-$5\arcsec$ from the center in a position angle of $-90^{\circ}$~deg at $v\sim5211$~km~s$^{-1}$ ($v-v_{\rm sys}\simeq-53$~km~s$^{-1}$).  A part of the feature is also detected at $v\sim5251$~km~s$^{-1}$ ($v-v_{\rm sys}\simeq-13$~km~s$^{-1}$, denoted E in Figure \ref{fig:CO_Chan}(d)).  No emission, except for the main disk emission, is detected at other velocity channels.  All the features from $v\sim5091$~km~s$^{-1}$ to $v\sim5251$~km~s$^{-1}$ seems to be hardly reproduced by a single stream of infalling gas since all the components are not aligned in the same position angle.   \cite{Lim2008} also pointed out that the inner filament and the western filament are separate structures because of their different kinematic structures, and we confirmed this in our high resolution images. 

\begin{figure*}
  \begin{center}
   \includegraphics[width=170mm]{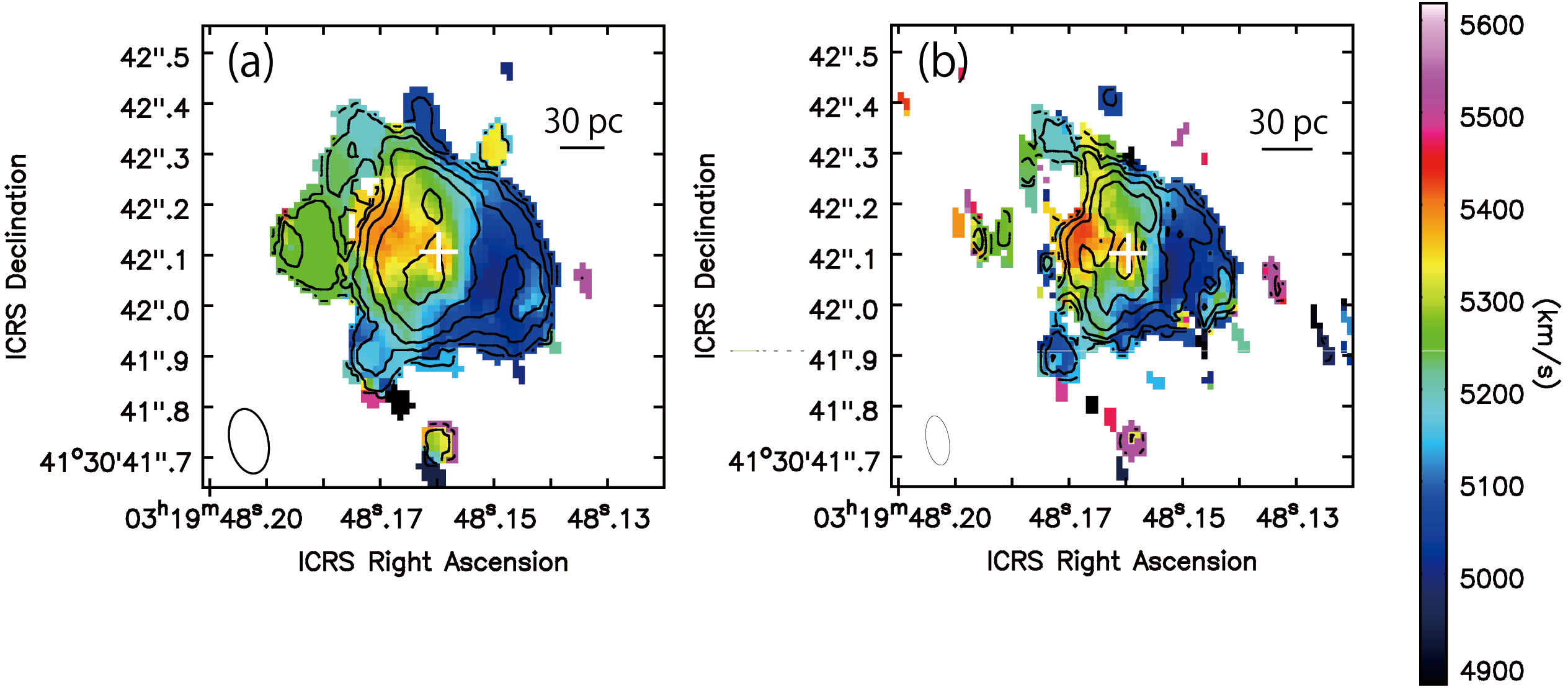}
 \end{center}  
\caption{(a) Contours of the naturally-weighted moment 0 map overlaid on the moment 1 map of HCN(3-2) in color.  The cross symbol indicates the AGN position, which was identified as the peak position of the continuum emission. (b) The same as (a), but the image is created using the Briggs weighting with the robust parameter of 0.5.  The images are plotted at the region where the flux density is greater than 5$\sigma$ ($1\sigma=0.95$~mJy for the naturally weighted image and $1\sigma=0.84$~mJy for the Briggs weighted image). The contours are plotted at the level of $20$~km~s$^{-1}\times3\sigma\times (-1, 1, 2, 4, \cdots, 512)$.  The image rms is measured within a circle with a radius of $1.8\arcsec$ (150 pixels) at the image center at line-free channels (averaged over 30-channels around $6000$~km~s$^{-1}$.).  The beam size is $0.129\arcsec \times 0.076\arcsec$ at the position angle of $12.1^{\circ}$ and $0.098\arcsec \times 0.045\arcsec$ at the position angle of $9.4^{\circ}$ for natural weighting and Briggs weighting, respectively.}
  \label{fig:HCN}
\end{figure*}

\begin{figure*}
  \begin{center}
   \includegraphics[width=170mm]{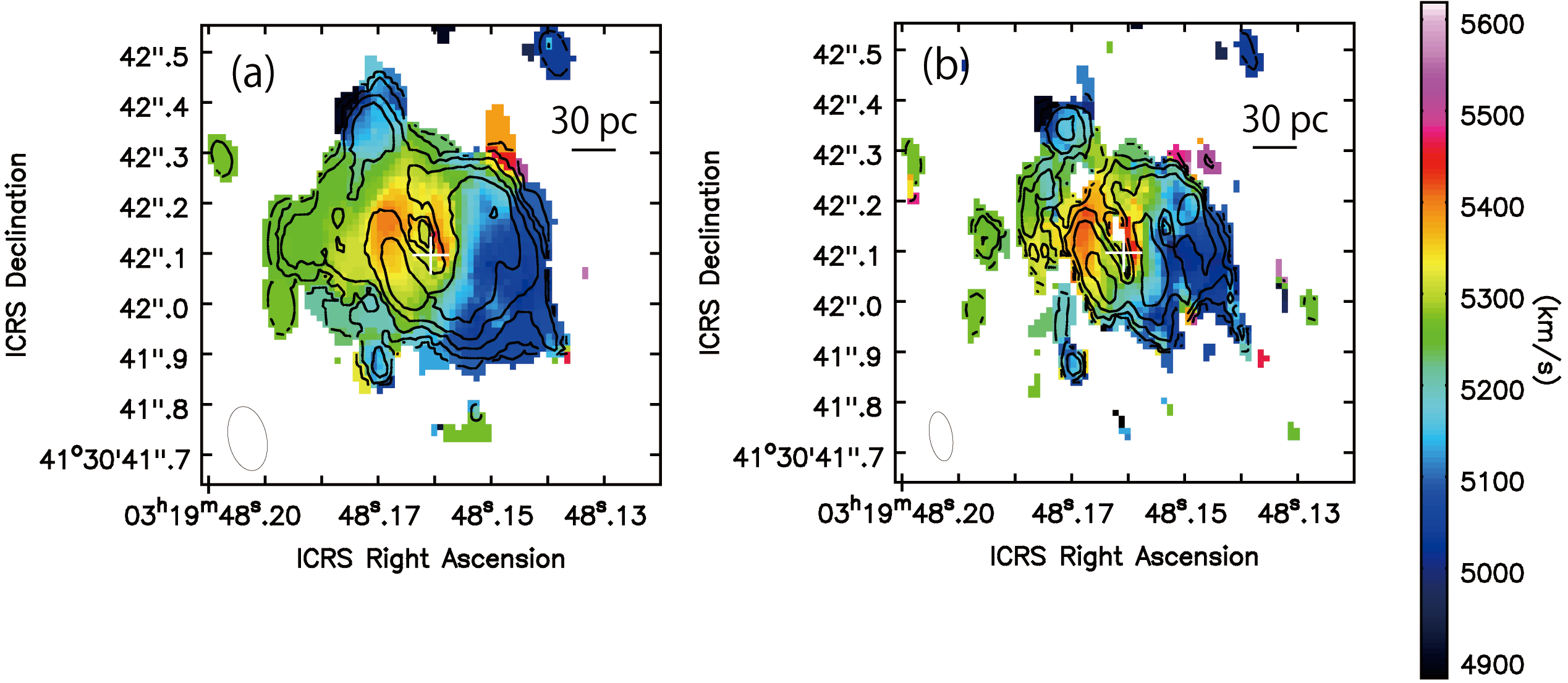}
 \end{center}  
\caption{(a) Contours of the naturally-weighted moment 0 map overlaid on the moment 1 map of HCO$^{+}$(3-2) in color.  The cross symbol indicates the AGN position, which was identified as the peak position of the continuum emission. (b) The same as (a), but the image is created using the Briggs weighting with the robust parameter of 0.5.  The images are plotted at the region where the flux density is greater than 5$\sigma$ ($1\sigma=1.05$~mJy for the naturally weighted image and $1\sigma=0.96$~mJy for the Briggs weighted image).  The contours are plotted at the level of $20$~km~s$^{-1}\times3\sigma\times (-1, 1, 2, 4, \cdots, 512)$.  The image rms is measured within a circle with a radius of $1.8\arcsec$ (150 pixels) at the image center at line-free channels (averaged over 30-channels around $6000$~km~s$^{-1}$.).  The beam size is $0.129\arcsec \times 0.076\arcsec$ at the position angle of $12.0^{\circ}$ and $0.099\arcsec \times 0.045\arcsec$ at the position angle of $9.3^{\circ}$ for natural weighting and Briggs weighting, respectively.}
  \label{fig:HCO}
\end{figure*}

\subsection{The HCN(3-2) and HCO$^{+}$(3-2) Lines}\label{sect:HCN-HCO}
\begin{figure}
\begin{center}
   \includegraphics[width=85mm]{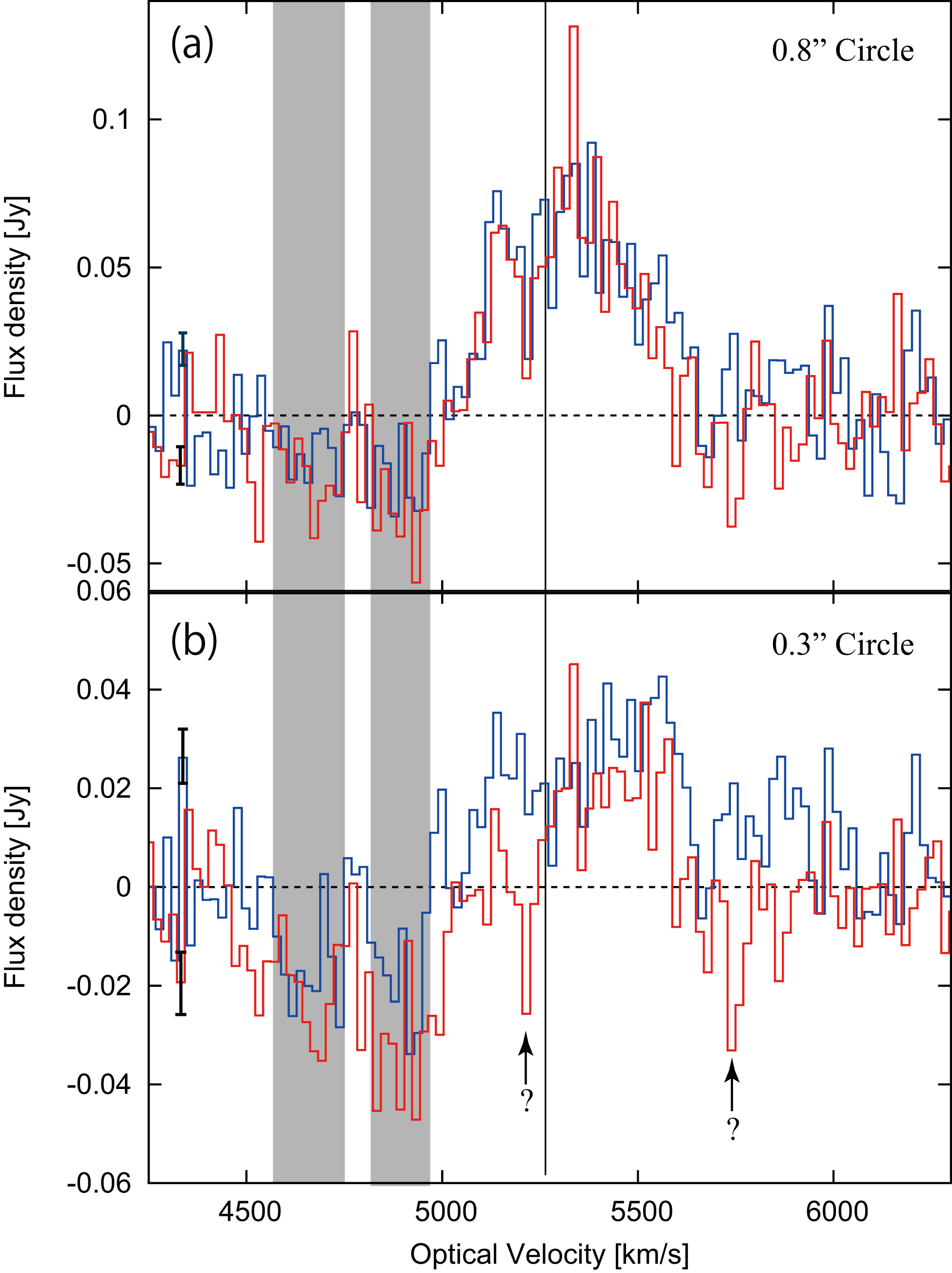}
\end{center}
 \caption{(a) The HCN(3-2) (blue) and HCO$^{+}$(3-2) (red) spectrum integrated over the region of $0.8\arcsec$-diameter circle at the center. (b) The same as (a), but the spectrum is computed over the region of $0.3\arcsec$-diameter circle at the center.  Error bars corresponding to $\pm1\sigma$ uncertainty are plotted in each panel (5.5~mJy for HCN(3-2) and 6.3~mJy for HCO$^{+}$(3-2)).  The error estimation is explained in the text (see \S\ref{sect:Obs}).  The vertical line indicates the systemic optical velocity of 5264~km~s$^{-1}$.  The velocity ranges where the absorption features are detected most significantly in both the HCN(3-2) and HCO$^{+}$(3-2) are highlighted by gray rectangles.  Additional deficits are seen in the HCO$^{+}$(3-2) spectrum, as indicated by arrows, but there is no corresponding deficits in the HCN(3-2) spectrum.}
  \label{fig:HCN-HCO_spectrum}
\end{figure}

Figure \ref{fig:HCN}(a) and Figure \ref{fig:HCO}(a) show the contours of naturally-weighted moment 0 maps overlaid on the moment 1 maps of the HCN(3-2) and HCO$^{+}$(3-2), respectively.  The images created with the Briggs weighting are also shown in Figure \ref{fig:HCN}(b) and Figure \ref{fig:HCO}(b).  These emissions are detected only within a radius of 100~pc, and the morphological and kinematical structures are also similar to those of the CO(2-1) emission.  Those molecules also show the velocity gradient, as seen in the CO(2-1) emission.  Figure \ref{fig:HCN-HCO_spectrum}(a) shows the spectrum integrated over the central $0.8\arcsec$.  Those spectral shapes are similar to the CO(2-1) spectrum, which is peaked at $\sim5330$~km~s$^{-1}$ and ranging from 5000~km~s$^{-1}$ to 5600~km~s$^{-1}$.  Previous single dish observations showed a consistent line width and line peak \citep{Salome2008, Bayet2011}. 

\begin{figure*}
\begin{minipage}{0.5\hsize}
  \begin{center}
   \includegraphics[width=85mm]{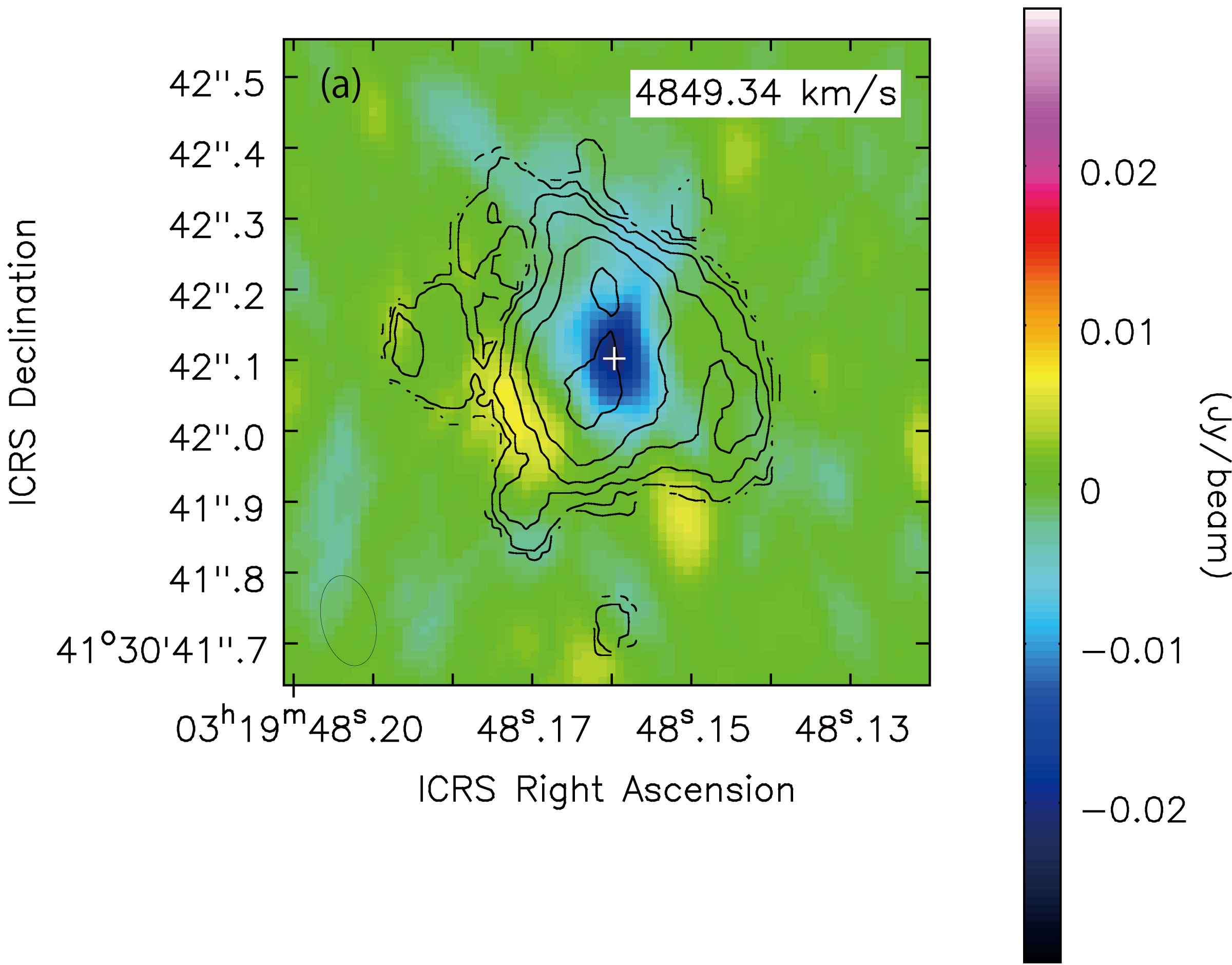}
 \end{center}  
\end{minipage}
\begin{minipage}{0.5\hsize}
\begin{center}
\includegraphics[width=85mm]{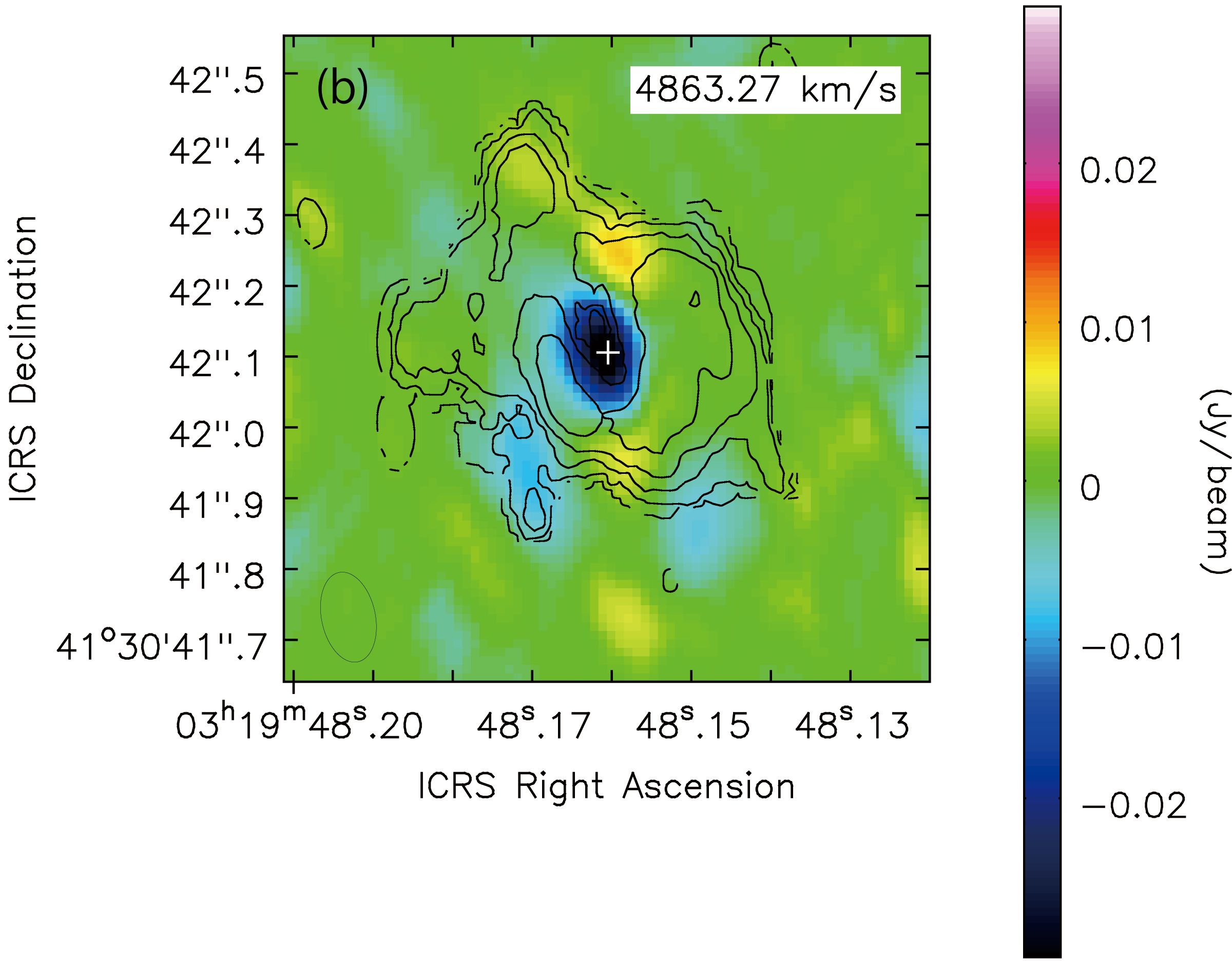}
\end{center}  
\end{minipage}
\caption{(a) Contours of the HCN(3-2) moment 0 map overlaid on the HCN(3-2) intensity map at $v=4849$~km~s$^{-1}$. (b) Contours of the HCO$^{+}$(3-2) moment 0 map overlaid on the HCN(3-2) intensity map at $v=4863$~km~s$^{-1}$.  Plus sign in both figures indicates the location of AGN.}
  \label{fig:HCN-HCO_absorption}
\end{figure*}

\begin{figure}
\begin{center}   
\includegraphics[width=85mm]{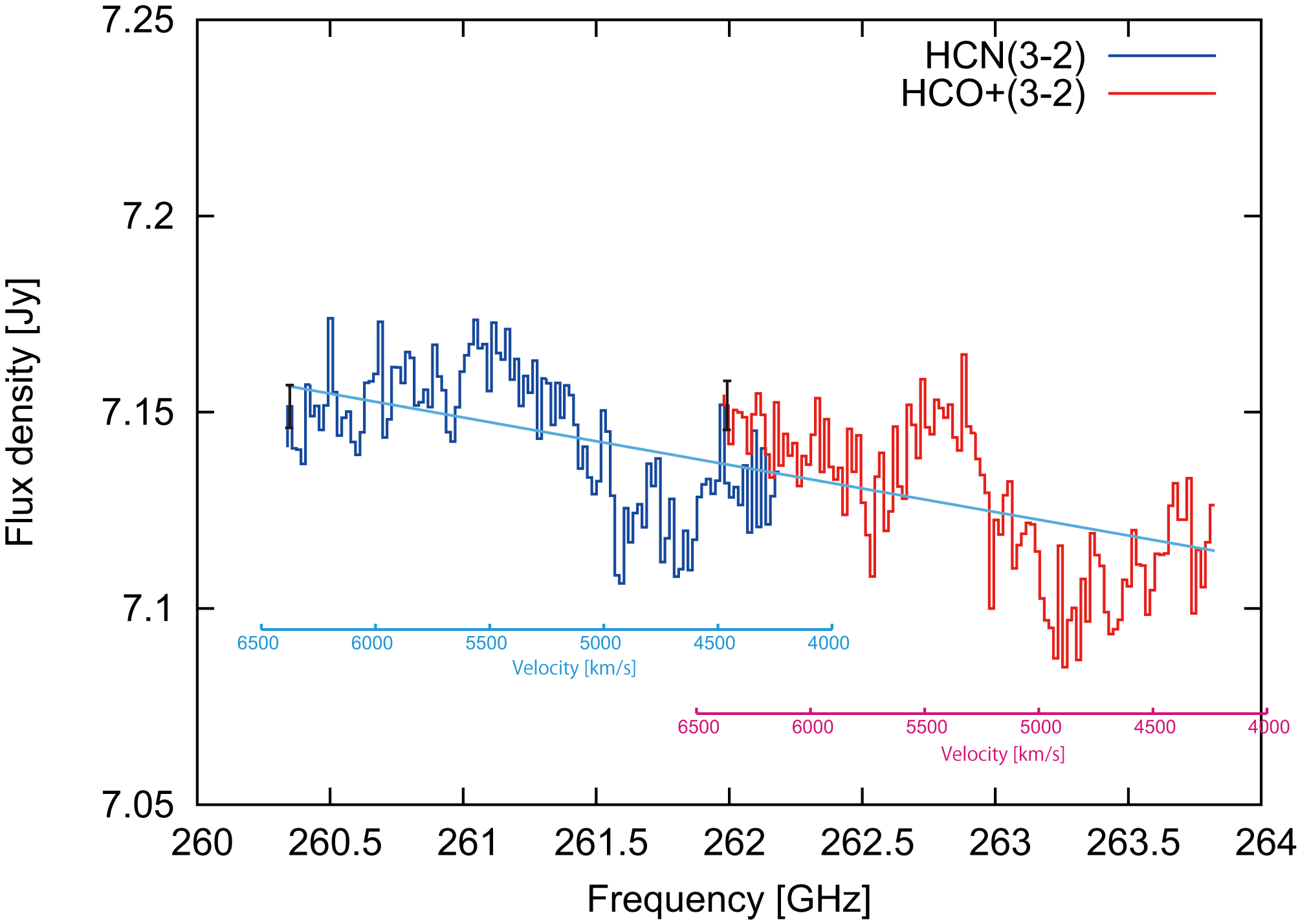}
\end{center}
 \caption{(a) The HCN(3-2) (blue) and HCO$^{+}$(3-2) (red) spectra measured on the continuum unsubtracted images.  The scales indicated in cyan and magenta correspond to the velocity for the HCN(3-2) and HCO$^{+}$(3-2) lines, respectively.  The flux was calculated within the ellipse ($0.15\arcsec\times0.21\arcsec$) slightly larger than the beam shape to cover all the emissions.  Uncertainties are plotted as $\pm1\sigma$.  Estimation of uncertainties is explained in the text (see \S\ref{sect:Obs}).  To highlight the spectral features, we also show the power-law fit indicated by the cyan line.  The fit was done using all spectral channels in both the HCN(3-2) and HCO$^{+}$(3-2) data.  The fitted power-law index is $-0.44$.  Note that there is a slight offset between the HCN(3-2) and HCO$^{+}$(3-2) data in the y-axis at the overlapping frequencies.  The difference is $\sim0.15$\% of the averaged flux density.  This is probably due to the absolute amplitude calibration error.}
  \label{fig:HCN-HCO_spectrum_nocontsub}
\end{figure}

One notable finding is that absorption features are detected at velocities of $\sim$4600-4900~km~s$^{-1}$ (Figure \ref{fig:HCN-HCO_spectrum}).  The significance is more than $3\sigma$ and $4\sigma$ at the deepest absorption channels for HCN(3-2) and HCO$^{+}$(3-2), respectively.  Since there is no corresponding major transitions of molecules at these frequencies, those features are most likely the blue-shifted components of the HCN(3-2) and HCO$^{+}$(3-2) lines shifted by 300-600~km~s$^{-1}$ with respect to the systemic velocity.  Figure \ref{fig:HCN-HCO_absorption} shows an example of a channel map where the absorption is detected.  This indicates that the absorption is caused on the direction to the central bright continuum associated with the AGN jet.  Flux density of the background continuum emission is about 7~Jy~beam$^{-1}$ while the peak flux of the absorption features is about 40~mJy~beam$^{-1}$, yielding the optical depth of 0.0057.  Under the assumption of the local thermodynamic equilibrium and a covering factor of 1, we obtain the H$_{2}$ column density of $2.3\times10^{22}$~cm$^{-2}$.  Here we assume a HCN-H$_{2}$ abundance ratio of $10^{-9}$ \citep{Smith2014}.  To check that these features are not an artifact caused by the process of continuum subtraction, we also show the spectrum measured on the continuum unsubtracted images in Figure \ref{fig:HCN-HCO_spectrum_nocontsub}.  Troughs in the spectrum are clearly seen between 261.6~GHz and 261.9~GHz for the HCN(3-2) and between 263.2~GHz and 263.7~GHz for the HCO$^{+}$(3-2).  

\begin{figure}
\begin{center}
   \includegraphics[width=85mm]{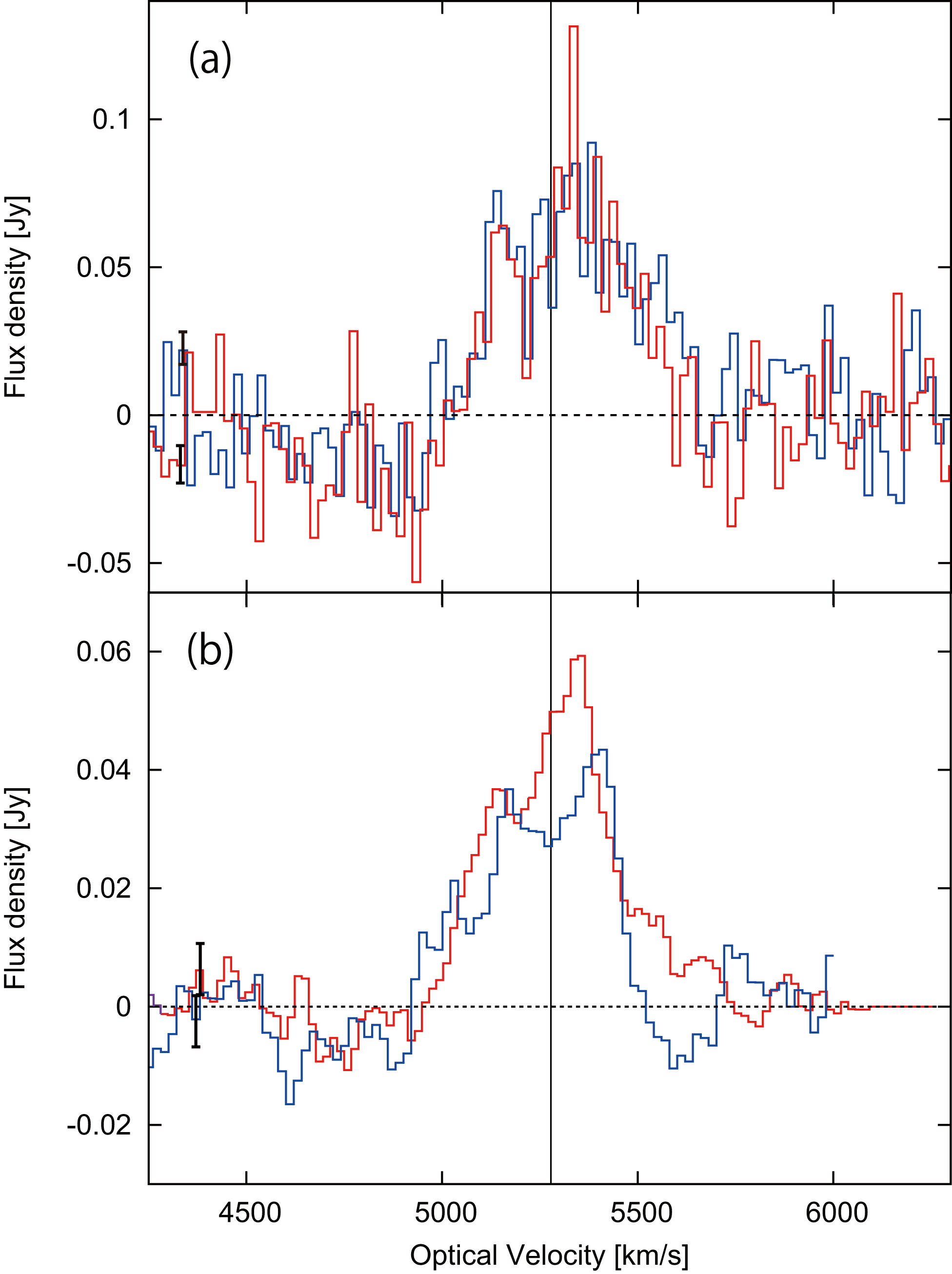}
\end{center}
 \caption{The comparison of the HCN(3-2) and HCO$^{+}$(3-2) spectra. (a) The same as Fig. \ref{fig:HCN-HCO_spectrum}(a). (b) The HCN(3-2) (red) and HCO$^{+}$(3-2) (blue) spectra of the archive data.  The vertical line indicates the systemic optical velocity of 5624~km~s$^{-1}$.  Uncertainties are plotted as $\pm1\sigma$ in the panel.  Estimation of uncertainties is explained in the text (see \S\ref{sect:Obs}).}
  \label{fig:HCN-HCO_spectrum_new-archive}
\end{figure}

To confirm the detection of absorption features, we also reduced archival data which observed NGC~1275 as the phase calibrator (Project ID: 2013.1.01102.S).  This data covered the frequency of the HCN(3-2) and HCO$^{+}$(3-2) lines.  Figure \ref{fig:HCN-HCO_spectrum_new-archive}(b) shows the spectrum of the those lines.  Absorptions in the spectrum could be marginally detected at the velocity of $\sim4620$-4900~km~s$^{-1}$ with the significance of at most $3\sigma$ for HCN(3-2) and $2\sigma$ for HCO$^{+}$(3-2) at the deepest absorption channels.  Low significance might be due to a relatively large beam size for the archival data.  However, we note that the marginal detection does not conflict with the detection with our data because of a possible time variation of the properties of absorbers.  We will discuss this in detail in section \ref{sect:absorption}.  

\section{DISCUSSION}
\subsection{Filaments}
Our ALMA observations detected two distinct filamentary structures -- the one is seen at $\sim5091$~km~s$^{-1}$ at a position angle of $\sim-80$~deg (A, B, and C in Figure \ref{fig:CO_Chan}), and another is seen at $v\sim5211$-5251~km~s$^{-1}$ in a position angle of $\sim-90$~deg (D and E in Figure \ref{fig:CO_Chan}).  Each component has a total H$_{2}$ gas mass of $\sim10^{6}~M_{\sun}$ (The assumptions used for this estimate is discussed in \S\ref{sect:InnerGasDisk}.).  Note that the estimated mass should be a lower limit since the total emissions cannot be recovered because of the lack of short interferometric baselines.  Lower velocity components mainly constitute the western filament while the higher velocity ones constitute the inner filament.   
Previous authors argued that the large-scale gas filaments can be infalling and settling into the CND \citep[e.g.][]{Scharwachter2013}.  However, these features may not be a representation of a single infalling stream since they are not simply connected to each other in terms of both morphology and kinematics.  We have not so far detected a direct evidence of the connection between the filaments and the CND.  

Yet, the complex filaments in NGC~1275 can be a natural consequence of CCA enhanced by thermal instabilities.  It is suggested by numerical simulations of elliptical galaxies that the growth of thermal instabilities in the hot halo/cluster gas leads to the condensation of cold clouds and filaments \citep{Gaspari2013, Gaspari2017}.  As a result, the accretion flow from cluster scales to the central black hole is dominated by CCA, i.e., the cold clouds show complex structures and non-ballistic orbits \citep{Pizzolato2010, Gaspari2013, Gaspari2017}.  The observed complex morphology of the CO filaments seems to be consistent with the simulations.  As we noted in section \ref{sect:intro}, deflection of jets and strong variation in the AGN luminosity of NGC~1275 is observed.  This is also consistent with the predictions of the CCA.  Our observations may not be sufficient to reconstruct the entire gas structures because of the lack of short interferometric baselines.  Observations with a compact antenna configuration would aid to image more diffuse and extended structures.   

How much of the cold gas is accreted by the SMBH?  We can make a crude estimate if the cold accretion is the dominant feeding mechanism for the NGC~1275 system.  Assuming each component is infalling to the center, we can roughly estimate the mass accretion rate as
\begin{eqnarray}
\dot{M}_{\rm cold}&=&\frac{M_{\rm cold}}{t_{\rm dyn}} \nonumber\\
&=& M_{\rm cold}\left( \frac{r}{v} \right)^{-1} \nonumber  \\
&=& 0.1\times\left[\frac{M_{\rm cold}}{10^{6}M_{\sun}}\right] \left[ \frac{r}{1~{\rm kpc}} \right]^{-1}\left[ \frac{v}{100~{\rm km~s}^{-1}} \right] M_{\sun}~{\rm yr}^{-1}.
\end{eqnarray}
Here $M_{\rm cold}$ is the cold gas mass, $v$ is the infalling velocity, and $r$ is the distance to each clump from the AGN.  For the first-order approximation, we assume that 3-dimensional velocity of each component is similar to the observed line-of-sight velocity.  As an example, the most pronounced component (component A in Figure \ref{fig:CO_Chan}(b) has a velocity of $\sim170$~km~s$^{-1}$ relative to the systemic velocity.  Each of components A-E in Figure \ref{fig:CO_Chan} has a H$_{2}$ gas mass of $\sim10^{6}M_{\sun}$ (see section \ref{sect:InnerGasDisk} about the details of conversion from the observed flux density to gas mass).  Taking into account that the estimated gas mass is a lower limit because of missing flux, we obtain $\dot{M}>0.1M_{\sun}$~yr$^{-1}$.  Similar estimates were done by \cite{Lim2008}, and they obtained a mass accretion rate of several tens of solar masses per year.  Thus, our estimate does not conflict with the estimate by \cite{Lim2008}.  \cite{Fujita2016} estimated the Bondi power of NGC~1275 using {\it Chandra} X-ray observations of the hot gas and the argument of momentum balance between the jet/cocoon and ambient medium.  The resultant Bondi accretion rate was $7.4\times10^{-5}M_{\sun}$~yr$^{-1}$ considering current jet activities (see Fujita et al. (2016) for more details).  This suggests that the cold accretion can be more efficient than the hot accretion on kpc scale of NGC~1275.

\subsection{Inner Gas Disk}{\label{sect:InnerGasDisk}}
Rotating disk-like morphology and kinematics are detected in the CO(2-1), HCN(3-2), and HCO$^{+}$(3-2) lines.  The gas disk spatially coincides with that of the warm H$_{2}$ gas disk \citep{Scharwachter2013}, and the velocity structure is also very similar between the two data sets.  Although only our data alone cannot give a robust constraint on the gas temperature, the temperature of the CND traced by the cold molecular lines (CO(2-1), HCN(3-2), and HCO$^{+}$(3-2)) is probably smaller than 100~K \citep[e.g.][]{Myers1978}.  On the other hand, the H$_{2}$ emission can trace the gas with the temperature of greater than 1000~K \citep{Wilman2005, Scharwachter2013}.  Therefore, it is natural to expect that the cold and warm gas phases are distributed at the same radii but are stratified.  This multiphase gas structure can be realized if the gas disk consists of many small gas clumps where the temperature of outer surface of clumps is raised by shocks or UV and X-ray irradiation so that the H$_{2}$ molecules can be collisionally excited while the temperature of the inner part remains at low temperature and is where the cold molecular lines are emitted.  Such a multiphase-gas system is also predicted by the CCA simulations \citep{Gaspari2017}.  

We calculate the total H$_{2}$ gas mass using equations (3) and (4) in \cite{Solomon2005} and obtain $\sim4\times10^{8}~M_{\sun}$.  Here we get the CO(2-1) integrated flux density of 40~Jy$\cdot$km~s$^{-1}$ within the central $1\arcsec$ with a velocity FWHM of 400~ km~s$^{-1}$.  We assumed the CO(2-1)-to-CO(1-0) ratio of one and the CO luminosity-to-H$_{2}$ mass conversion factor in our Galaxy ($\alpha_{\rm CO}=4.8~M_{\sun}$~pc$^{-2}$(K~km~s$^{-1}$)$^{-1}$) \citep{Solomon2005}.  The derived gas mass is about 20\% of the total gas mass of the inner filament presented in \cite{Lim2008}.  If we use $\alpha_{\rm CO}=$0.6-0.8, which is widely used as the conversion factor for gas-rich galaxies at high redshift and local U/LIRGs \citep{Downes1998, Papadopoulos2012}, we obtain the total gas mass of $\sim$5-7$\times10^{7} M_{\sun}$.  \cite{Izumi2016} reported a correlation between the mass of cold molecular gas ($M_{\rm gas}$) in the CND and black hole accretion rate $\dot{M}$ for Seyfert galaxies.  Using the derived correlation relation and $M_{\rm gas}=4\times10^{8}~M_{\sun}$, we obtain $\dot{M}_{\rm BH}$ of $\sim5\times10^{-2}M_{\sun}$~yr$^{-1}$, which shows an agreement with the accretion rate ($\sim1\times10^{-1}M_{\sun}$~yr$^{-1}$) inferred from the bolometric luminosity \citep{Nagai2017} within a factor of two.

The CCA simulations predict that the condensation evolution in a quiescent rotating hot halo leads to the formation of a multiphase disk \citep{Gaspari2017}.  In the equatorial plane ($x$-$y$ plane), the infalling gas spins-up increasing the rotational velocity to conserve the initial angular momentum.  The gas also gradually goes into free fall along the $z$ direction (perpendicular to the equatorial plane).  This drives a gas motion following a conical helix, but after a long duration (40~Myr in \cite{Gaspari2017}), the angular momentum in $x$ and $y$ directions are canceled out by the cloud-cloud collisions and the multiphase gas converges to the equatorial disk.  Low gas density along the jet direction is indeed suggested by VLBI observations \citep{Fujita2016, Fujita2017}.  Both cold gas (CO(2-1), HCN(3-2) and HCO$^{+}$(3-2)) disk and the warm H$_{2}$ disk extend to a radius of $\sim0.3\arcsec$ ($\sim100$~pc).  We can conclude that the rotational motion dominates the turbulent motion at $\sim100$~pc in terms of the CCA.  \cite{Gaspari2017} reported that the dense disk blocks the inflow of gas at large distance from the center, leading to the expansion of the disk radius with time.  Larger cold gas disks, extending to a few kpc, are found in many other early-type galaxies by the CARMA CO(1-0) imaging survey of early-type galaxies \citep{Alatalo2013} and radio galaxies by ALMA CO(2-1) \citep{Ruffa2019}.  As compared to those early-type galaxies or radio galaxies, the gas disk in NGC~1275 is remarkably small.  This might be explained by relatively strong jet activity.  NGC~1275 has repeated strong AGN outbursts, and thus could have uplifted the gas and stirred up the turbulence by the jet/cocoon in the polar region \citep{Churazov2001, Ensslin2006}.  This can suppress gas infall from the hot atmosphere.  The extended emission line nebula around NGC~1275 \citep{Hatch2006, Fabian2008, Gendron-Marsolais2018} could be a signature of such uplifted gas by the radio jets/bubbles.  All galaxies in \cite{Alatalo2013} shows much lower radio continuum flux than NGC~1275, so the jet feedback can be less efficient for those galaxies.  This scenario also supports the lack of kpc-scale gas disks in PKS~0745-191 \citep{Russell2016}, Phoenix cluster \citep{Russell2017}, and A2597 \citep{Tremblay2016} for which all these sources show strong jet activities.  
Contrary, radio galaxies Hydra~A and 3C~31 have a kpc-scale molecular gas/dust disks \citep{Okuda2005, Fujita2013, Hamer2014}.  In particular, Hydra A shares many properties, including the strong jet activity, with NGC~1275 \citep{Hamer2014}.  The existence of a cold gas disk on kpc scales of Hydra~A may suggest that the cluster gas is less turbulent for some reasons so the CCA cannot develop.  Low turbulence is also indicated by low velocity dispersion of the CO(2-1) gas revealed by the ALMA observations \citep{Rose2019}.  The future X-ray satellite mission XRISM would help to uncover the gas motion of BCGs in more detail.

A velocity gradient of the gas is observed in position angle of $\sim70\degr$.  This rotation axis is nearly parallel to the innermost jet axis which was revealed by the space VLBI observation with Radioastron \citep{Giovannini2018}.  \cite{Hiura2018} reported the non-linear jet/hotspot motion on parsec-scale and discussed the possibility that the non-linear motion was caused by the precession of the jet base.  Nevertheless, the precession angle is only a few degrees.  Thus, the disk rotation axis is still approximately consistent with the jet axis.  The coincidence of the disk and jet axis alignment probably indicates that the molecular gas disk is connected to the inner accretion disk, which is responsible for the jet launching. 
Alternatively, this might indicate that the spin axis of the central SMBH is aligned to the molecular gas disk if the Blandford-Znajek mechanism \citep{Blandford1977} is responsible for the jet production.

\begin{figure*}
\begin{center}
\includegraphics[width=18cm]{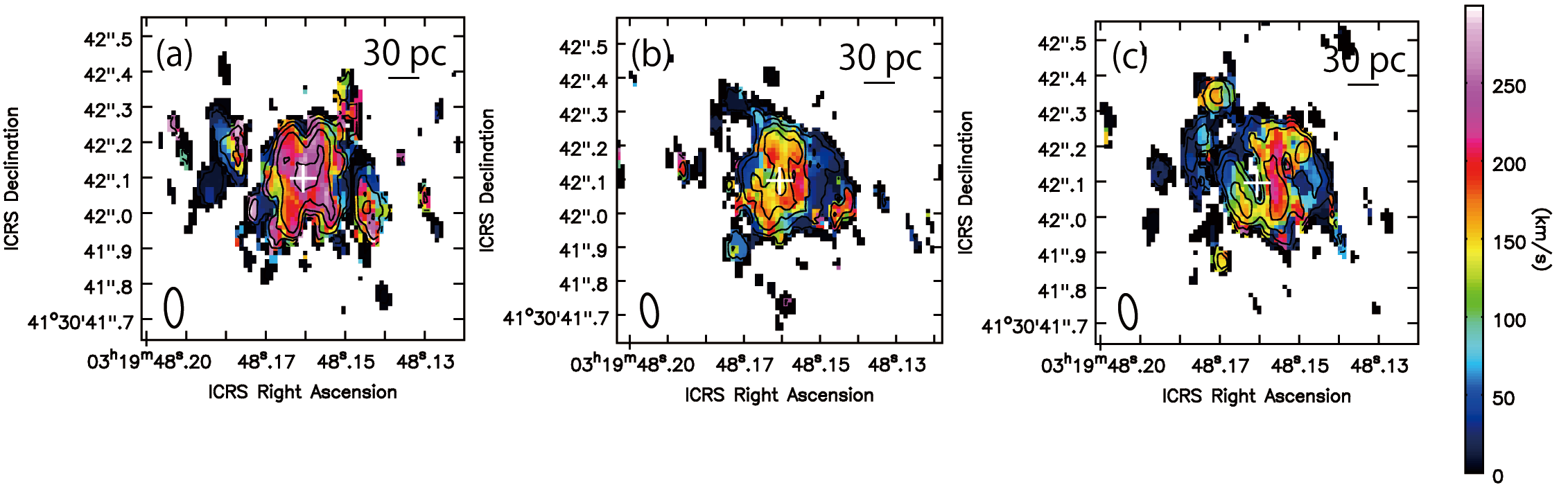}
\end{center}
\caption{Contours of moment 0 map overlaid on colors of moment 2 map for (a) CO(2-1), (b) HCN(3-2), and (c) HCO$^{+}$(3-2).  The maps are made with the Briggs weighting of robust parameter of 0.5.  Contour levels of (a), (b), and (c) are the same for Fig. \ref{fig:CO}(b), Fig. \ref{fig:HCN}(b), and Fig. \ref{fig:HCO}(b), respectively.}
\label{fig:mom2}
\end{figure*}

Let us estimate the accretion timescale of the molecular gas in the observed disk.  If we assume that the cold gas disk is thin and the kinetic viscosity is the main source of angular momentum transfer, the accretion timescale ($\tau$) can be given by $\tau=r^{2}/(\alpha c_{s} h)= r^{2}/(vh)$\citep[e.g.][]{Shakura1973, Pringle1981}, where $r$ is the disk radius, $\alpha$ is the viscosity parameter, $c_{s}$ is the sound speed, $v$ is the turbulent velocity, and $h$ is the scale height.  Assuming a thin disk ($h=0.1r$), we get
\begin{equation}
\tau\sim4\times10^{7}\times\left[\frac{r}{\rm 100~pc}\right]\left[\frac{v}{\rm25~km~s^{-1}}\right]^{-1} {\rm yr}.
\end{equation}
Here we adopt the observed typical velocity dispersion (see Figure \ref{fig:mom2}) for the turbulent velocity ($v$).  We should notice that the velocity dispersion at the inner part of CND in Figure \ref{fig:mom2} does not reflect true turbulent velocity.  An apparent large velocity dispersion ($>100$~km~s$^{-1}$) is probably an artifact which is caused by convolving the synthesized beam to a large velocity gradient (see $-0.05\arcsec\leq x \leq0.05\arcsec$ in Figure \ref{fig:CO_PVD}).  We therefore adopt typical velocity dispersion in the outer region ($v\sim25$~km~s$^{-1}$).  We can roughly estimate the averaged accretion rate ($\dot{M}_{\rm gas}$) of the gas disk within a radius of 100~pc as $\dot{M}_{\rm gas}\simeq M_{\rm gas}/\tau$.  For $M_{\rm gas}=5$-$7\times10^{7}~M_{\odot}$ and $M_{\rm gas}=4\times10^{8}~M_{\odot}$ (see \S\ref{sect:InnerGasDisk}), we get $\dot{M}_{\rm gas}=1$-$2~M_{\odot}$~yr$^{-1}$ and $\dot{M}_{\rm gas}=10~M_{\odot}$~yr$^{-1}$, respectively.  This accretion rate of cold gas is again higher than Bondi accretion rate of $7.4\times10^{-5} M_{\odot}$\citep{Fujita2016}.  This manifests that the cold accretion is a dominant channel on tens-pc scale.    

Comparison between accretion timescale and cooling timescale of hot ICM would give a hint for the timescale of AGN activity.  \cite{Sanders2004} showed the cooling time of hot gas in NGC~1275 at the radius down to 5~kpc. Within 10~kpc, cooling time is estimated to be $5\times10^{8}$~yr and does not change very much as a function of distance.  The density profile of BCGs is well represented by a $\beta$-model with a flat density profile within the core radius.  The flat density profile is favored even in the inner region ($r\sim100$~pc) of NGC~1275 \citep{Fujita2016}.  Since the cooling time is inversely proportional to the density, it should be reasonable to assume that the cooling timescale does not change much within the core radius.  We here use the cooling time of $10^{8}$~yr on 100-pc scale.  This is longer than the accretion timescale.  If the angular momentum of the black hole accretion disk is controlled by the cold gas of the CND, the jet-launching direction can be also controlled by the global angular momentum of the CND. We therefore expect that the jet direction would be kept more or less in the same direction on the accretion timescale ($\sim10^{7}$~yr).  However, the subsequent mass accretion from larger spatial scales may change the global angular momentum in the case of CCA, and thus the jet launching direction can change from the current direction on longer timescales.  This idea is supported by the observed radio morphology:  While the radio axis within 1~arcsec ($<1$~kpc) is similar to the current jet axis \citep{Taylor1996, Silver1998, Walker2000, Asada2006}, the radio lobes/bubbles on kpc scales are misaligned from the inner radio axis \citep{Pedlar1990}.  The change in the jet direction, as well as spreading energy over large volumes by buoyant bubbles \citep{Zhang2018}, helps to heat up the galactic halo or cluster atmosphere efficiently and globally \citep{Cielo2018}, which can mitigate the issue that radio-mode feedback tends to be efficient only in the jet direction.  

Even though our observations have not reached to the spatial scale of the putative torus for the AGN unification scheme ($\sim$1-10~pc), the observed molecular gas (in particular HCN(3-2) and HCO$^{+}$(3-2)) shows a central concentration.  It seems to be reasonable to assume that the molecular gas disk is reaching to the pc scale.  As we discussed in section \ref{sect:HCN-HCO}, we detected the absorption lines of HCN(3-2) and HCO$^{+}$(3-2) only in blue-shifted velocity ranges.  The absence of a redshifted absorption feature suggests that the molecular gas disk at pc scale, where the gas is likely infalling to the AGN, should not intercept our line of sight toward the AGN.  In order to realize this, the molecular gas disk must be thin ($h<r$ where $h$ is the disk height and $r$ is the disk radis) since we are viewing the AGN from a relatively large angle ($65\degr\pm15\degr$: \cite{Fujita2017}).  From the AGN unification viewpoint, NGC~1275 was originally classified as a Narrow Line Radio Galaxy or Seyfert 2, which required obscuration of the central engine by the putative torus.  However, the broad H$\alpha$ wing was later identified and NGC~1275 was classified as a Seyfert 1.5/LINER \citep{Veron2006, Sosa-Brito2001}. The broad H$\beta$ emission with the velocity FWHM of 4150-6000~km~s$^{-1}$ as well as the Pa$\alpha$ and C\,{\footnotesize IV} has been also recently detected \citep{Punsly2018}.  Those detections of broad emissions lines are consistent with the presence of a thin disk as suggested from our observations.  The thin disk was also reported in another low-luminosity AGN, NGC~1097 \citep{Izumi2017}.  \cite{Izumi2017} suggested that a thinner disk in NGC~1097 than in NGC~1068 \citep{Imanishi2018} arose as the result of its low luminosity ($L_{\rm bol}=8.6\times10^{41}$~erg~s$^{-1}$; \cite{Nemmen2006}), which is supported by theoretical predictions \citep[e.g.][]{Elitzur2006}.  It is not clear if the relatively high luminosity of NGC~1275 ($L_{\rm bol}=4\times10^{44}$~erg~s$^{-1}$; \cite{Levinson1995}) can be explained by the same mechanism.  We need more samples to improve our understanding of the thickness of the disk/torus.  

\cite{Hitomi2018} has recently reported the detection of Fe-K$\alpha$ line emission associated with the AGN of NGC~1275.  The observed velocity width of Fe-K$\alpha$ is $\sim500$-1600~km~s$^{-1}$.  This velocity width suggests that the fluorescing matter is located at a distance of $\sim$1.4-14 pc from the black hole, if we attribute it to the Keplerian rotation velocity of a black hole mass $8\times10^{8}~M_{\sun}$.  The absence of time variability of the emission over 10 years by {\it Hitomi}, {\it XMM-Newton}, and {\it Chandra} also supports the inferred distance of the fluorescing matter.  The small equivalent width of the Fe-K$\alpha$ line suggests that the fluorescing matter of NGC~1275 is gas with a relatively low-column density and/or low-covering fraction as compared to other Seyfert galaxies.  Hitomi Collaboration et al. concluded that the source of the Fe-K$\alpha$ line is a low-covering-fraction molecular torus or a rotating molecular disk which probably extends from a parsec to hundreds of parsecs scale.  An intriguing comparison is that the velocity width of the HCN(3-2) and HCO$^{+}$(3-2) does not exceed $\sim600$~km~s$^{-1}$ while the velocity width of Fe-K$\alpha$ is $\sim500$-1600~km~s$^{-1}$.  A part of the fluorescing matter can be our observed gas disk, but the higher velocity side of the Fe-K$\alpha$ emission can originate in other sources, perhaps the inner part of the disk, e.g, the $[$F\,{\footnotesize II}$]$ emitters \citep{Scharwachter2013} or free-free absorber \citep{Walker2000, Fujita2017}.

\subsection{Origin of Absorption Feature}\label{sect:absorption}
The absorption feature is detected at $v\sim$4600-4900~km~s$^{-1}$ in both HCN(3-2) and HCO$^{+}$(3-2).  The absorbers must be located somewhere between the observer and the AGN of NGC~1275.  The absorption against the continuum flux is at most $\sim40$~mJy~beam$^{-1}$ (or $\sim0.6$\% attenuation), which corresponds to an optical depth of 0.0057.  The total H$_{2}$ column density of the absorbers is calculated to about $2.3\times10^{22}$~cm$^{-2}$ using the equation (2) given by \cite{Sanhueza2012}.  Here we assume the excitation temperature of 100~K and the HCN-to-H$_{2}$ conversion factor of $10^{-9}$.  We adopt this conversion factor based on the results from the abundance ratio estimates in the galactic circumnuclear disk \citep[(0.2-4.1)$\times10^{-9}$;][]{Smith2014} and dense molecular outflow in Mrk~231 \citep[$0.5\times10^{-9}$;][]{Aalto2015}.  

What is the origin of the absorbers?  The background continuum emission is mostly coming from the jet of $\sim1.2$-pc length \citep{Nagai2010, Nagai2014} while the extended radio emission arises from larger scale from 10~pc to 10~kpc \citep{Romney1995, Silver1998, Pedlar1990, Asada2006}, which are probably associated with the past episodes of jet activities, are negligibly small at this frequency \citep[e.g.,][]{Silver1998}.  Thus, the absorbers are likely located between the observers and the innermost jet with projected size of about 1.2~pc.  The beam filling factor of the jet is about $10^{-3}$.  Thus, it is very unlikely that a gas clump at a large distance from the AGN causes the absorptions.  It is natural to expect that the absorptions are associated with the clouds close to the nucleus.  Since the absorbers are blue-shifted from the systemic velocity by $\sim300$-$600$~km~s$^{-1}$, they must be moving toward us.    

One promising candidate is outflowing materials associated with the approaching side of the radio jet.  The approaching jet is moving toward the south, and the total extension is about 1.2~pc \citep{Nagai2017}.  \cite{Nagai2017} and \cite{Kino2018} found the abrupt change in the position of hotspot of the southern jet in 2015-2016 with an enhancement of polarized emission.  This hotspot motion, as well as the enhancement of the polarization, can be caused by the interaction between the jet and a dense ambient medium, probably a dense gas clump with the size of $\sim0.1$~pc.  The dense cloud can be expelled from the AGN by this interaction \citep[e.g.][]{Wagner2012}.  We speculate that the outflowing absorbers are candidates for the dense cloud.  Such a jet-ISM interaction was also identified in the compact radio jet of 4C~12.50 as HI absorption by VLBI observations \citep{Morganti2013}.  Assuming a spherical clump with the radius of 0.1~pc, we obtain the total gas mass of the cloud $\sim3 M_{\sun}$ and the H$_{2}$ gas density $\sim7.5\times10^{4}$~cm$^{-3}$.  This gas density is in agreement with that estimated by the Faraday rotation observations under the equipartition condition within a factor of 4 \citep{Nagai2017} and the one estimated by the hotspot dynamics within its uncertainty \citep{Kino2018}.  This gas mass can be a lower limit if the clump has been elongated by the jet interaction, as seen in the 4C~12.50 case.  Alternatively, the absorber could be the molecular outflow expelled from the innermost region of the CND by the radiation pressure from the AGN.  The observed velocity ($\sim300$-$600$~km~s$^{-1}$) is in fact consistent with that predicted by numerical simulations \citep{Wada2015}.  Further observations with other lines may help to better understand the origin of absorber.  In particular, the shock tracer (e.g., SiO) would be detected if the absorber is associated with the jet-ISM interaction.

Previous VLBI observations \citep{Nagai2017, Kino2018} suggest that the interaction between the jet and the cloud started in late 2015.  The archival data shown in section \ref{sect:HCN-HCO} was taken in 2015 June when the jet interaction had not started or just started.  Therefore, it can be a natural consequence that the absorption lines are less obvious in the archival data.

One may ask why the absorption feature is not detected for the CO(2-1) line.  Although the detailed physical mechanism is not clear, several studies have argued that the harsh environment in the close vicinity of the AGN and the influence by the radio jet can increase the HCN or HCO$^{+}$ transitions.  \cite{Meijerink2011} modeled the intensity ratio of molecular species under the condition of strong cosmic-ray and mechanical heating.  For the condition of strong mechanical heating, the HCN/CO intensity ratio increases, and this trend can be boosted, depending on the cosmic-ray heating rate.  \cite{Harada2013} also reported that the intensity ratio of HCN/CO can be larger than unity in the region where the X-ray irradiation is strong.  This might explain the absence of CO(2-1) absorption lines since the dense cloud could be mechanically heated by the jet interaction and also be strongly irradiated by cosmic-rays and X-rays from the AGN.  The enhancement of HCN and HCO$^{+}$ was also reported in the outflow of Mrk~231\citep{Aalto2015} and NGC~253 \citep{Walter2017}.  One intriguing comparison is that \cite{Michiyama2018} reported the detection of the HCN(1-0) and HCO$^{+}$(1-0) emission in the outflow from the southern nucleus of the merging galaxy NGC~3256 where a low-luminosity AGN is located.  On the other hand, the HCN(1-0) and HCO$^{+}$(1-0) are undetected in the outflow from the northern nucleus where there is no AGN signature.  They claimed that the enhancement of the HCN(1-0) and HCO$^{+}$(1-0) emissions is caused by the formation of dense clumps associated with the jet-ISM interaction, possibly via shocks.

The main disk emission shows a double-peak structure in the HCO$^{+}$(3-2) spectrum for both our data and archive data (see Figure \ref{fig:HCN-HCO_spectrum} and Figure \ref{fig:HCN-HCO_spectrum_new-archive}).  This spectral shape was not observed in the previous single dish observations \citep{Bayet2011}.  The deficit between the two peaks in the HCO$^{+}$ spectrum is possibly caused by the absorption, which is evident in the spectrum only at the image center where the AGN is located (see Figure \ref{fig:HCN-HCO_spectrum}(b)).  Another absorption-like feature may be seen in the redshift-side at $v\sim5750$~km~s$^{-1}$.  However, these two features are not seen in the HCN(3-2) spectrum.  We therefore need a confirmation by new observations.

\section{Black Hole Mass}
SMBH mass is a fundamental parameter to determine the AGN physics in connection with the galaxy evolution.  It also determines the Schwarzschild radius, which characterizes various physical quantities, such as Bondi radius and SoI (SoI$=GM_{\rm BH}/\sigma_{v}^2$ where $G$ is the gravitational constant, $M_{\rm BH}$ is the black hole mass, and $\sigma_{v}$ is the central stellar velocity dispersion.).  Although there is a well-known empirical correlation between the SMBH mass and stellar velocity dispersion ($M_{\rm BH}-\sigma$ relation, e.g., \cite{Bettoni2003, Kormendy2013, Bosch2016}), dynamical measurement of the SMBH mass within the SoI, where the gravitational potential is thought to be dominant, is important to derive a robust SMBH mass.  Using the stellar dispersion velocity of $250$~km~s$^{-1}$ and $M_{\rm BH}$ of $10^{8.61}M_{\sun}$ \citep{Bettoni2003}, we obtain SoI of 60~pc.  Therefore, molecular gas dynamics revealed by our ALMA observations should be useful to give a robust constraint on the SMBH mass of NGC~1275.  Previously, the SMBH mass measurements using the warm H$_{2}$ gas dynamics were made by \cite{Wilman2005} ($3\times10^{8}M_{\odot}$) and \cite{Scharwachter2013} ($8\times10^{8}M_{\odot}$).  Although those measurements probed the gas dynamics with a similar spatial resolution with our data, the cold gas (CO(2-1), HCN(3-2), and HCO$^{+}$(3-2)) and warm H$_{2}$ gas can be stratified as we discussed in section \ref{sect:InnerGasDisk}.  Therefore, an independent SMBH mass measurement is still intriguing.

We here model the rotating molecular gas disk traced by the CO(2-1) and directly measure the SMBH mass by following a standard procedure as described in \citet{Onishi2017, Davis2017, Davis2018}.
We use an image cube with natural weighting (see Section~\ref{sect:Obs}), that yields a beam size of $0.144\arcsec \times 0.077\arcsec$.
The fitting area is defined here as the central $1.55\arcsec \times1.55\arcsec$ region and from $4687$ to $5580$ km~s$^{-1}$, thus $129\times129\times47$ pixels. 
We model the galaxy mass profile (stars and the SMBH) by using the Multi Gaussian Expansion method \citep[{\tt MGE};][]{Emsellem1994, Cappellari2002}, and include gas mass self gravity by using the publicly available \textsc{KINematic Molecular Simulation} \citep[{\tt KinMS}; ][]{Davis2013} tool\footnote{\url{https://github.com/TimothyADavis/KinMS}}.

\begin{figure}
\begin{center}
\includegraphics[width=7.5cm]{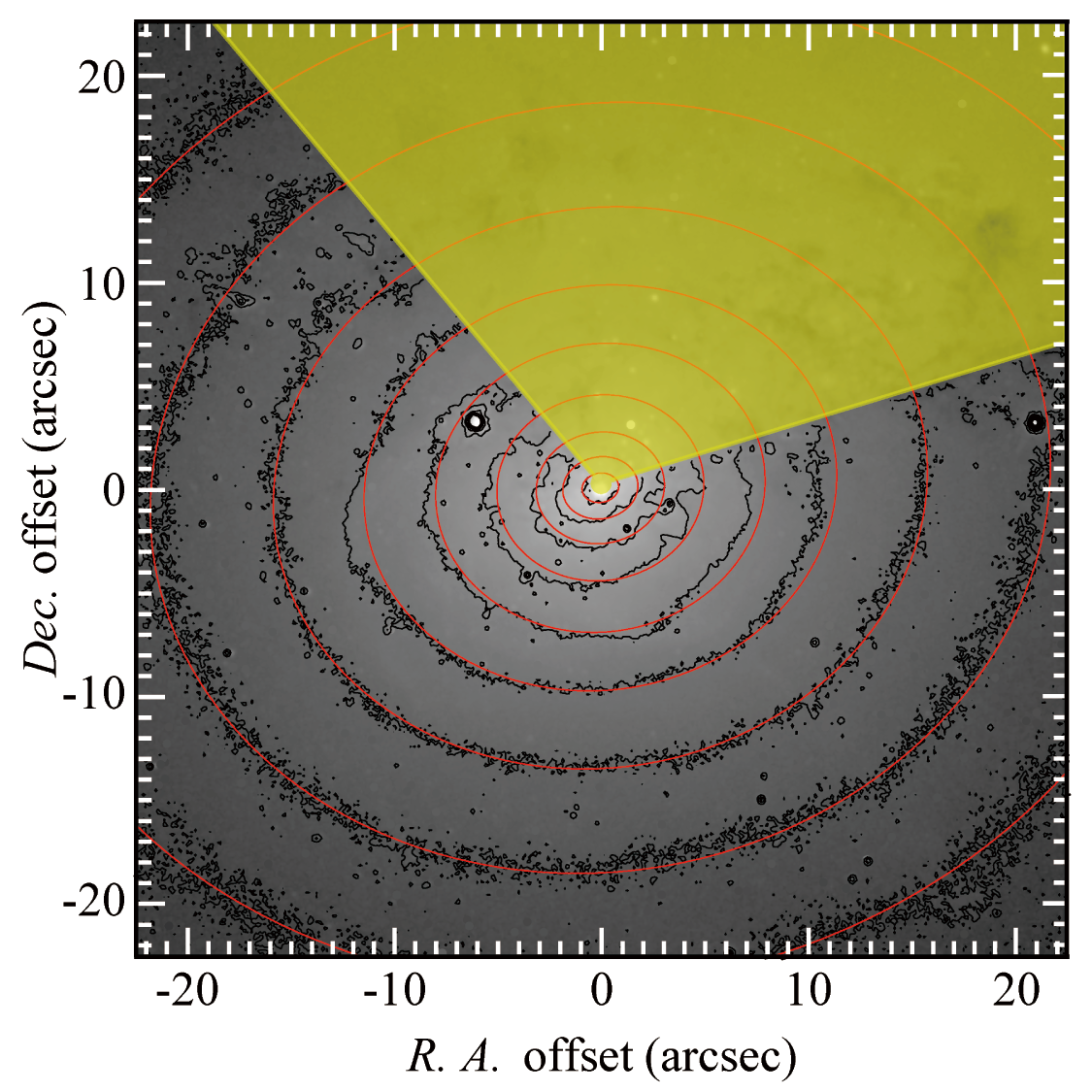}
\end{center}
\caption{A Multi Gaussian Expansion (MGE) model (red contours) overlaid to a {\it HST} $I$-band archival image (F814W filter on Wide Field Planetary Camera 2).  We exclude regions highlighted in yellow so to avoid any effect from the AGN (a region within $0.5\arcsec$ radius from the AGN) and from dust attenuation (cone-shaped region in the north west.)}
\label{fig:HSTMGE}
\end{figure}

We use {\it HST} $I$-band archival image taken with the F814W filter on Wide Field Planetary Camera 2 to model the stellar mass component. 
The {\it HST} PSF for this observation was estimated by using TinyTim Version~6.3 \citep[][]{Krist2011}.
Before modeling the $I$-band luminosity distribution with multiple Gaussians, we mask a region within $0.5\arcsec$ radius from the AGN position, comparable to the size of the PSF (FWHM of $\sim0.8\arcsec$), to avoid any light from the AGN. 
The dust-attenuated region in the north west is also masked (see Figure~\ref{fig:HSTMGE}).
We then take the unmasked region of the image ($100\arcsec\times100\arcsec$) and fit the luminosity distribution by using the PSF and the procedure {\tt mge\_fit\_sectors\_regularized}, contained in the {\tt MGE\_fit\_sectors} IDL package\footnote{http://purl.org/cappellari/software} of \citet{Cappellari2002}.
Figure~\ref{fig:HSTMGE} shows the {\it HST} $I$-band image (greyscale) with the mask (colored in yellow) and the MGE model (red contours) overlaid.
A spatially uniform stellar mass-to-light ratio ($M/L$) is then multiplied by the model luminosity profile to create a stellar mass model.
The SMBH mass (delta function) is then added at the center.
NGC 1275 is known to have a substantial amount of gas at the center, that possibly affects the kinematics \citep[][]{Scharwachter2013}.
The molecular gas mass of $4\times10^{8}M_{\sun}$, estimated in Section~\ref{sect:InnerGasDisk}, is thus included in the mass model.
We also consider a different estimation of the molecular gas mass ($5\times10^{7}M_{\sun}$, also described in Section~\ref{sect:InnerGasDisk}) in a different fitting run in order to see the effect.

By using the mass model described above with parameters of $M/L_{I}$ and the SMBH mass, a circular velocity curve is calculated from the {\tt MGE\_circular\_velocity} procedure contained in the {\tt JAM} package of \citet{Cappellari2008}, assuming an axisymmetric potential and circular motion.
By utilizing the {\tt KinMS}, we take observational effects such as beam smearing and velocity resolution into account, and generate dynamical models of the molecular gas disk.
The molecular gas disk is assumed to be in a form of an exponential disk ($\exp{-r}$/$r_{\rm scale}$); where $r$ corresponds to radius and $r_{\rm scale}$ is the scale length).
We consider inclination angle, kinematic position angle, $r_{\rm scale}$, gas velocity dispersion (assumed to be uniform over the disk), central position (R.A., Dec, and velocity) and the disk luminosity scaling as parameters to describe the disk.
For clarification, gas velocity dispersion given here represents the intrinsic velocity dispersion of gas, such as that occurred by turbulence.
These parameters determine each dynamical model of the molecular gas disk, that is created in the form of a simulated data cube. The simulated cube is then compared to the observed data to evaluate the difference.

We first searched for a set of parameters to minimize the difference between the observed data and the model by using {\tt mpfit} that is developed in {\tt IDL}.
The results from two different runs with $M_{\rm gas}=4\times10^{8}~M_{\sun}$ and $M_{\rm gas}=5\times10^{7}~M_{\sun}$ are summarized in Table~\ref{table:SMBH_param}.

\begin{figure}
\begin{center}
\includegraphics[width=7.5cm]{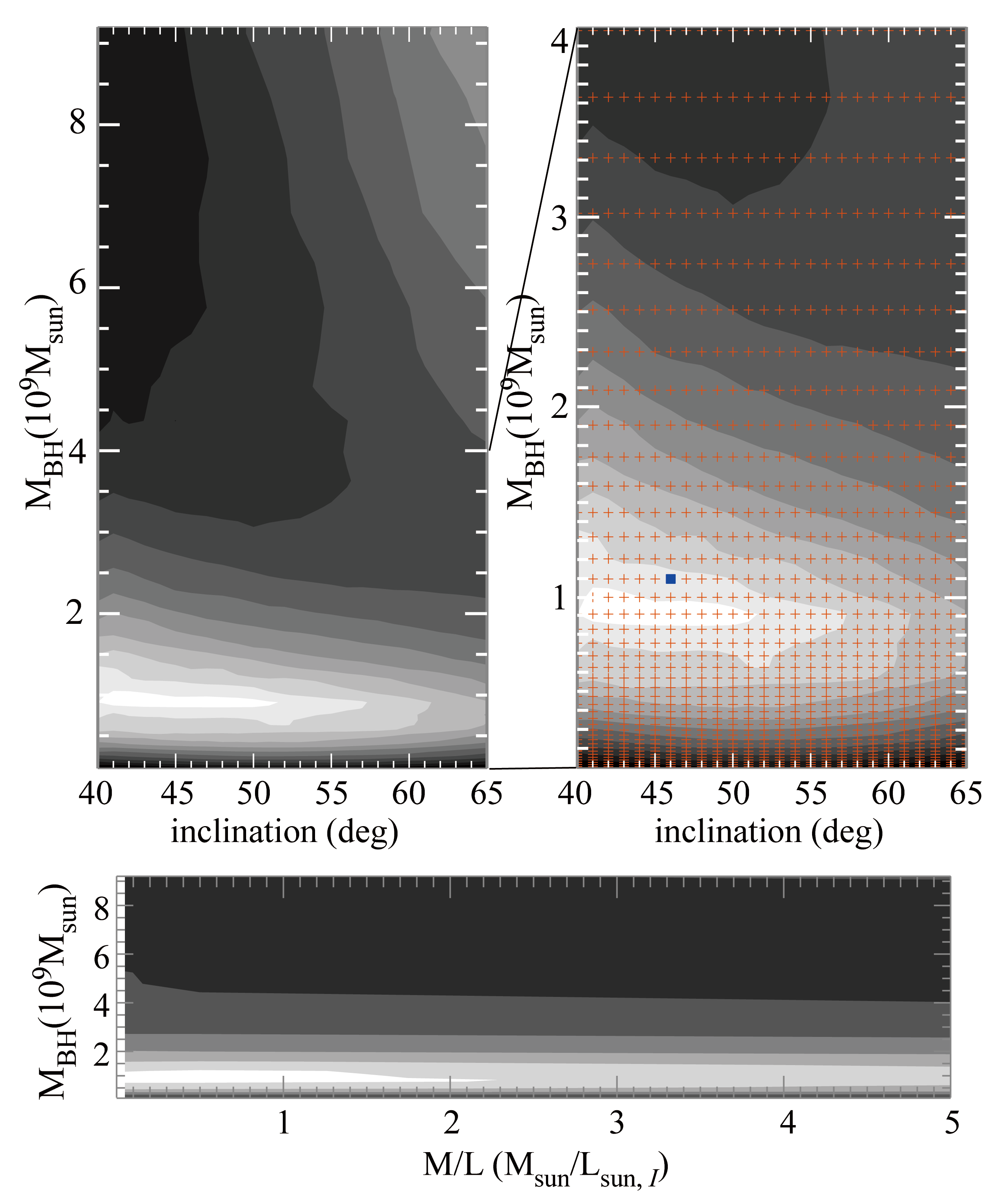}
\end{center}
\caption{\textit{Upper panels:}
Residual distribution in a parameter space of inclination ($40$ -- $65$
deg) and SMBH mass ($1\times10^{8}M_{\sun}$ -- $10\times10^{9}M_{\sun}$).
Here we assume the molecular gas mass of $4\times10^{8}M_{\sun}$ (see
Section~\ref{sect:InnerGasDisk}).
Left panel shows a full parameter space, and the right panel is zoomed into
$M_{\rm BH}=1\times10^{8}M_{\sun}$ -- $4\times10^{9}M_{\sun}$.
The best-fitting parameters obtained by using {\tt mpfit} are highlighted
in blue square.
Contours are spaced by $1$, $4$, $9$, $16$, $25$, $36$, $49$, $64$, $81$,
$100$ plus the minimum residual value reached in this grid search.
\textit{Lower panel:}
Residual distribution in a parameter space of $M/L$ ($0.05$ -- $5.00$
$M_{\sun}/L_{\sun, I}$) and SMBH mass ($1\times10^{8}M_{\sun}$ --
$10\times10^{9}M_{\sun}$).
This is to confirm that the $M/L$ do not affect the best-fitting SMBH mass.
}
\label{fig:SMBH_param_conts_mol4e8}
\end{figure}

\begin{figure}
\begin{center}
\includegraphics[width=7.5cm]{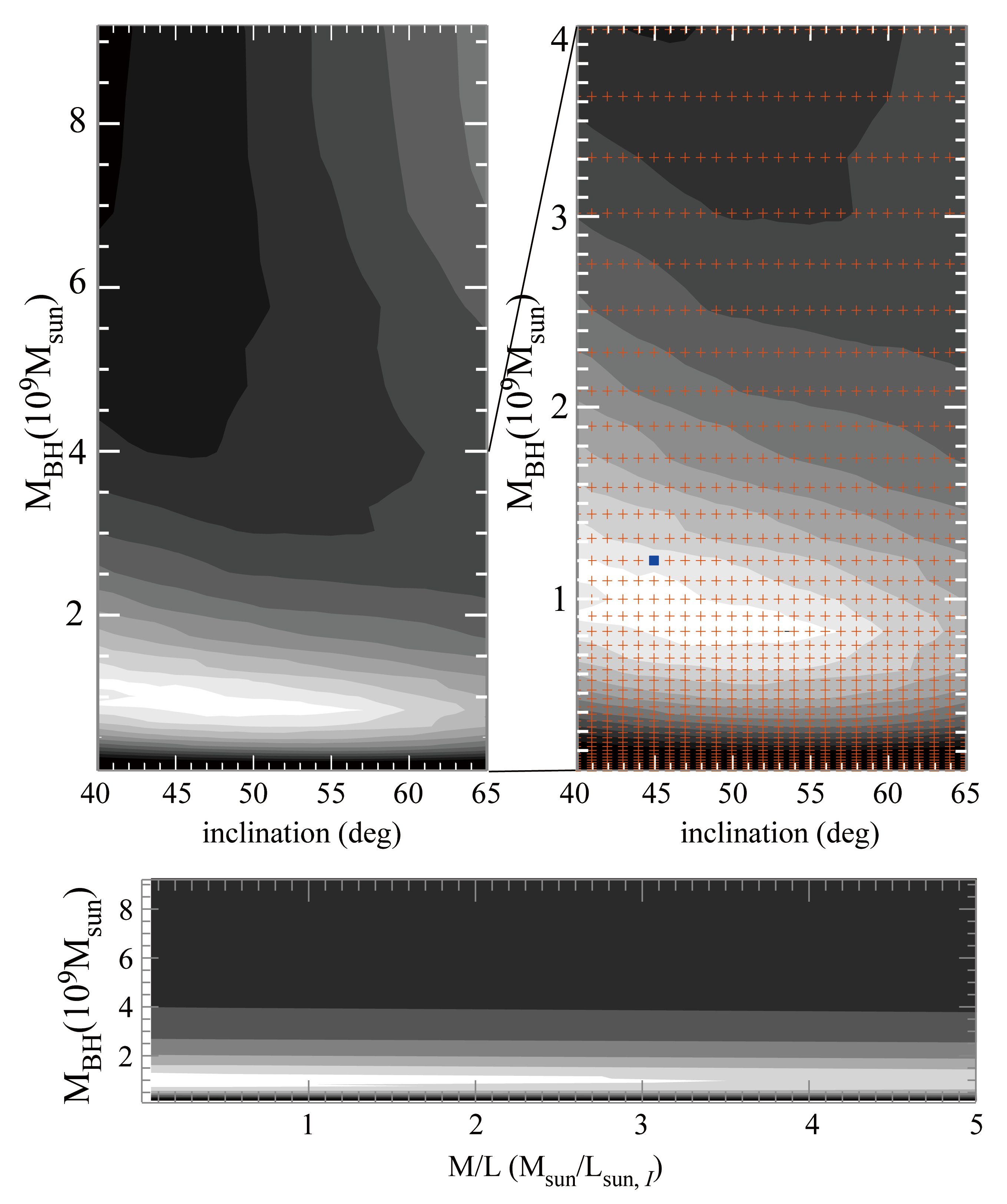}
\end{center}
\caption{Same as Figure~\ref{fig:SMBH_param_conts_mol4e8}, but the molecular gas mass is this time assumed to be $5\times10^{7}M_{\sun}$ (see Section~\ref{sect:InnerGasDisk}). }
\label{fig:SMBH_param_conts_mol5e7}
\end{figure}

We then evaluate error budgets for parameters those have crucial effect on the SMBH mass.
We here select gas disk inclination and the SMBH mass to search the parameters in grids, while other parameters are fixed to the {\tt mpfit} results.
$M/L_{I}$ does not affect the SMBH mass for the case of NGC~1275, suggesting that the stellar mass within the fitting area is ignorable \citep[][]{Scharwachter2013}. We confirm this by giving three different $M/L_{I}$ values, $0.05$, $0.50$ and $5.00$ to see negligible difference (see the lowest panels in Figures~\ref{fig:SMBH_param_conts_mol4e8} and \ref{fig:SMBH_param_conts_mol5e7}).
Figure~\ref{fig:SMBH_param_conts_mol4e8} (the case of $M_{\rm gas}=4\times10^{8}~M_{\sun}$) and Figure~\ref{fig:SMBH_param_conts_mol5e7} (the case of $M_{\rm gas}=5\times10^{7}~M_{\sun}$) show residual contours for the SMBH mass and inclination.
The residual is calculated in the same manner as the $\chi^{2}$, thus $({\rm observed} - {\rm model})^{2}/({\rm rms_{obs}})^{2}$.
This residual is useful to judge the fitting results but is not equivalent to the $\chi^{2}$, as the neighboring pixels in the observed cube are correlated.

\begin{figure*}
\begin{center}
\includegraphics[width=18cm]{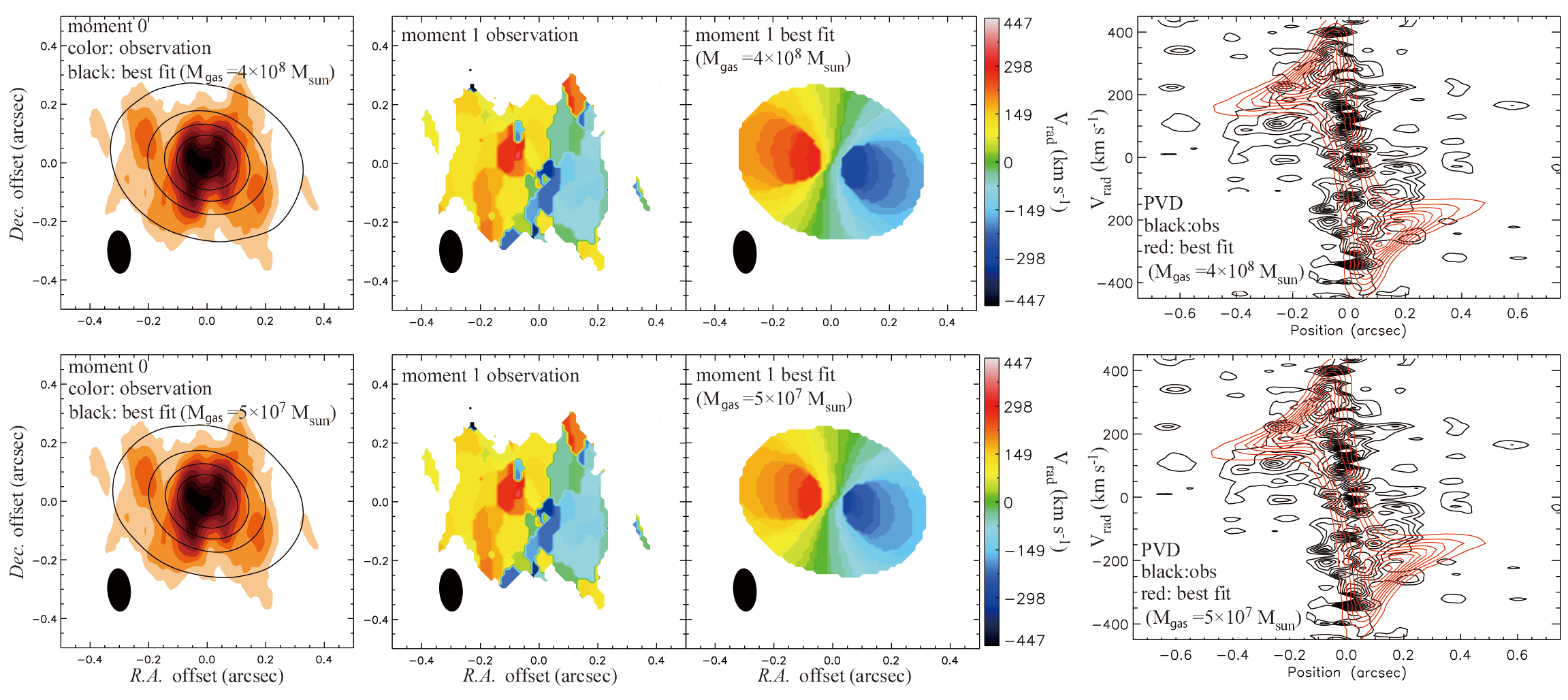}
\end{center}
\caption{\textit{Upper panels:} Comparison between the observation and the best-fit model with a molecular gas mass of $4\times10^{8}M_{\sun}$.
Colors are the observed moment 0 (leftmost), moment 1 (middle) and position-velocity diagram (rightmost) along the kinematic major axis (position angle of $75$~deg). Red contours in the position-velocity diagram represent the best-fit model.
\textit{Lower panels:} The same as the upper panels but the best-fit model is derived with a molecular gas mass of $5\times10^{7}M_{\sun}$.
See Table~\ref{table:SMBH_param} for the given parameter values.}
\label{fig:SMBH_bestfitmom01pvd}
\end{figure*}

The best fit values given by the {\tt mpfit} are $M_{\rm BH}=1.1\times10^{9}~M_{\sun}$, inclination of $46$~deg\footnote{Inclination angle of $0\degr$ and $90\degr$ correspond to the face-on disk and edge-on disk, respectively.}, and the $M/L_{I}=0.050M_\sun/L_{\sun, I}$ when molecular gas mass is $4\times10^{8}M_{\sun}$,
and $M_{\rm BH}=1.2\times10^{9}~M_{\sun}$, inclination of $45$~deg, and the $M/L_{I}=0.050~M_\sun/L_{\sun, I}$ when molecular gas mass is $5\times10^{7}~M_{\sun}$.
Comparisons with what we observed in a form of moment 0 and moment 1 and position-velocity diagram is shown in Figure~\ref{fig:SMBH_bestfitmom01pvd}.
Our assumption of exponential disk may not perfectly describe the molecular gas distribution of NGC~1275, but we consider the model distribution has only ignorable effect on the resulting SMBH mass.

The error budget is defined for the SMBH mass and inclination using the residual contours as given in Figures~\ref{fig:SMBH_param_conts_mol4e8} and \ref{fig:SMBH_param_conts_mol5e7}.
The large number of pixels ($129\times129\times47 \equiv N_{\rm dof}$; number of degree of freedom excluding negligible number of free parameters), however, gives non-negligible standard deviation to the $\chi^{2}$, as noted in \cite{vandenBosch2009} and \cite{Mitzkus2017}.
In order to avoid unrealistically small errors resulting from this $\chi^{2}$, we follow the approach of \citet{vandenBosch2009} to conservatively increase the $\Delta\chi^{2}$ level by multiplying all observed rms noise by $(2N_{\rm dof})^{1/4}$, that is exactly equivalent to decreasing the $\chi^{2}$ by $\sqrt{2N_{\rm dof}}$, and thus redefining the confidence level of $\Delta\chi^{2}$.
Again, the rescaled residual values are not equivalent to the $\chi^{2}$, but we judge the error by allowing all parameters that realize residuals less than $9+\chi_{\rm min}^{2}$ where $\chi_{\rm min}$ is the minimum rescaled residual value during the grid search ($940.6$ when $M_{\rm gas}=4\times10^{8}~M_{\sun}$, $926.6$ when  $M_{\rm gas}=5\times10^{7}~M_{\sun}$).
We obtained the best-fit SMBH mass of $M_{\rm BH}=(1.1 ^{+0.3}_{-0.5})\times10^{9}M_{\sun}$ and inclination of $46^{+11}_{-6}$~deg when molecular gas mass is $4\times10^{8}M_{\sun}$,
and $M_{\rm BH}=(1.2^{+0.1}_{-0.6}) \times10^{9}M_{\sun}$ and inclination of $45^{+14}_{-5}$~deg when molecular gas mass is $5\times10^{7}M_{\sun}$.
We here note that the inclination angle hits the lower limit of the search range, which is given from the stellar mass {\tt MGE} model.
The inclination will at least not reduce the SMBH mass down to zero, simply judging from the residual contours (Figures~\ref{fig:SMBH_param_conts_mol4e8} and \ref{fig:SMBH_param_conts_mol5e7}).

Since the estimated gas mass could be a lower limit because of the lack of short baselines, let us check how the missing mass affects the BH mass estimate. We assume that the whole structure of disk consists of our detected component with a radius of $\sim0.3\arcsec$ and a more extended component which is completely filter out in our observations.  We assume that the total gas mass of the inner filament of $1.8\times10^{9}~M_{\sun}$ \citep{Lim2008} uniformly distributes over the beam area of the observations by Lim et al. ($3\arcsec\times2.7\arcsec$).  The missing mass within a radius of $0.3\arcsec$ can be estimated as $(0.3\arcsec\times0.3\arcsec)\times(1.8\times10^{9})/(3\arcsec\times2.7\arcsec)=2\times10^{7}~M_{\sun}$.  This is very well within the SMBH mass error, and thus we conclude the missing mass will not give a big change to our result.

While the derived SMBH mass shows a good agreement with that derived from the warm H$_{2}$ gas dynamics \citep{Scharwachter2013}, there is a huge discrepancy with the SMBH mass using the virial estimate of broad emission lines ($\sim10^{7.5} M_{\sun}$, \cite{Onori2017, Koss2017}).  More recently, \cite{Punsly2018} also pointed out that the virial mass estimate can be 10-100 times smaller than the required SMBH mass in order to explain the bolometric luminosity.  If we take $M_{\rm BH}=10^{7.5}M_{\sun}$ and emission line velocity ($v_{\rm BLR}$) of $4500$~km~s$^{-1}$, the emission region radius ($r_{\rm BLR}$) is estimated to be $\sim0.007$~pc using the virial relation $r_{\rm BLR}=GM_{\rm BH}/v_{\rm BLR}^{2}$.  \cite{Netzer1993} suggested that the outer edge of the broad line region (BLR) conincided with the dust sublimation radius of the torus/CND, and reverberation mapping observations also show that the BLR extends out to the dusty torus \citep{Suganuma2006}.  Using equation (1) in \cite{Kishimoto2007} and UV luminosity estimated of $5\times10^{42}$~erg~s$^{-1}$ by \cite{Kino2017}, the dust sublimation radius can be estimated as
\begin{eqnarray}
\left[\frac{r_{\rm sub}}{\rm pc}\right]&=&0.029\left[\frac{L}{5\times10^{42}~{\rm erg~s}^{-1}}\right]^{0.5} \nonumber \\
&&\quad \times \left[\frac{T}{1500~{\rm K}}\right]^{-2.8}\left[\frac{a}{0.05~\mu{\rm m}}\right]^{-0.5}.
\end{eqnarray}
Here $L$ is the UV luminosity, $T$ is the sublimation temperature, and $a$ is the grain radius.  This estimate suggests that there is a discrepancy between $r_{\rm sub}$ and $r_{\rm BLR}$ by a factor of 4.  On the other hand, if we take $M_{\rm BH}=1\times10^{9}M_{\sun}$, we then get $r_{\rm BLR}$=0.2~pc, which is more than one order of magnitude larger than $r_{\rm sub}$.  \cite{Punsly2018} defined the deprojection factor of the orbital radius of the BLR by taking into account the BL~Lac nature of NGC~1275.  Assuming a small jet viewing angle, they derived $M_{\rm BH}=1\times10^{9}M_{\sun}$ from the virial mass estimate using the broad emission lines.  However, the small viewing angle assumption is not consistent with the viewing angle estimate using jet/counter-jet length ratio \citep{Fujita2017} unless the jet angle drastically changes from the jet base to parsec scale.  In conclusion, it is difficult to relate the SMBH mass estimate using broad emission lines with the observed nature of NGC~1275.  Since NGC~1275 exhibits repeated AGN outbursts, the ionizing photon flux from the accretion disk is probably highly time variable.  This could make the BLR complex, resulting in the discrepancies described above.

We here note that our model still stands on some assumptions that can be explored more carefully.
For example, the inclination angle and $M/L_{I}$ can have a radial dependence (such as would be the case for a warped disk), and the gas distribution can be better described with multiple Gaussians rather than the exponential disk.
Also, additional uncertainties can arise from the imperfectly known distance to the object. 
We can calculate the proper $\chi^{2}$ by comparing the data and the model in the {\it (u, v)} plane.
We however do not probe this in detail and leave these points for our future work.

\begin{table*}
\caption{Search range and best-fit values for model parameters}
\begin{center}
    \begin{tabular}{lccccc} \hline\hline
 & &\multicolumn{2}{c}{mol. gas $4\times10^{8}M_\odot$} & \multicolumn{2}{c}{mol. gas $5\times10^{7}M_\odot$}
\\ \hline
parameter & Search range &  \textsc{MPFIT} best & grid search error & \textsc{MPFIT} best & grid search error \\ \hline
SMBH ($M_\odot$)                                             & $10^{8}$ -- $10^{10}$ &  $1.1\times10^{9}$  &  $+0.3$, $-0.5$             & $1.2\times10^{9}$& $\pm3\times10^{8}$ \\
$M/L_{I}$ ($M_\odot/L_{\odot, I}$)                 & $0.01$ -- $1.00$             &  $0.050$                     &  --                                &  $0.050$                  & -- \\
incl. (deg)                                                             & $40$ -- $65$                   &  $46$                          &  $+11$, $-6$                     &$45$                   & $+14$, $-5$ \\
PA (deg)                                                                & $64.0$ --$80.0$                    &  $74.9$                     & --                               & $75.0$                  & -- \\
$r_{\rm scale}$ (arcsec)                                      & $0.05$ -- $0.20$                &  $0.12$                      & --                             & $0.12$                     & -- \\
gas $\sigma$ (km~s$^{-1}$)                               & $0$ -- $200$                  &  $22$                            &  --                                & $23$                    & -- \\
central offset x (arcsec)                                        & $-0.2$ -- $0.2$                &  $0.0$                           & --                            & $0.0$                      & -- \\
\phantom{central offset} y \phantom{(arcsec)}  & $-0.2$ -- $0.2$                & $0.0$                            & --                            & $0.0$                      & -- \\
\phantom{central offset} v (km~s$^{-1}$)         & $-30$ -- $30$                  & $0$                            & --                               & $0$                         & -- \\
luminosity scaling                                                & $20$--$50$                     & $28.0$                       & --                               & $27.4$                     & -- \\
\hline
\end{tabular}
\end{center}
\tablecomments{Parameter search region (column 1) and best-fit values (column 2 and 4) that minimizes the $\chi^{2}$ given from the \textsc{mpfit} procedure, ran for two different molecular gas mass $4\times10^{8}M_{\odot}$ and $5\times10^{7}M_{\odot}$.
Error bars are estimated from a grid search of three parameters (inclination, SMBH mass and stellar M/L; see text) and listed in the table (column 3 and 5).
See also Figure~\ref{fig:SMBH_param_conts_mol4e8} and \ref{fig:SMBH_param_conts_mol5e7} for the residual contours.}
\label{table:SMBH_param}
\end{table*}

\section{Conclusion}
We reported the ALMA observations of the CO(2-1), HCN(3-2), and HCO$^{+}$(3-2) lines on NGC~1275.  All three lines are detected within the central 100~pc significantly. Only the CO(2-1) emission is also marginally detected on the sub-kpc scale, though it is heavily resolved.  The sub-kpc CO(2-1) emission forms filamentary structures, but those cannot be represented by a single infalling stream both morphologically and kinematically.  Observed complex behavior of CO(2-1) is consistent with the chaotic cold accretion predicted by recent numerical simulations.  The mass accretion rate inferred from the CO gas can be higher than the Bondi accretion rate, which suggest that the cold accretion can be a dominant mechanism in this system. 

The main CO(2-1), HCN(3-2), and HCO$^{+}$(3-2) emissions appear to show disk morphology.  The velocity distribution shows a velocity gradient in a position angle of $\sim70^{\circ}$, which can be interpreted as the disk rotation.  The observed disk morphology and kinematics are similar to those of warm H$_{2}$ gas disk shown by previous Gemini observations.  This suggests that the cold and warm gas phases are distributed at the same radii but are stratified.  Intriguingly, the disk rotation axis is approximately the same with the jet axis on the subpc scale.  This may indicate that the cold gas disk is physically connected to the further inner accretion disk, which is responsible for jet launching, and thus the cold accretion may play an important role to characterize the AGN activity.  Higher spatial resolution would be crucial to study the nature of cold accretion in further inner region, as well as the connection with the FFA disk identified by VLBI observations.  

Overall, our observations suggest that the cold accretion can play an essential role for the AGN feeding in the representative of nearby BCGs, which are thought to be the most massive galaxies and the hosts of the most massive SMBH in the local Universe.  A better angular resolution with more extended ALMA configurations will allow us to resolve the Bondi radius ($\sim10$~pc) of NGC~1275 in future Cycles, which is crucial to better understand the physics of black hole accretion. 

We detected the blue-shifted absorption lines in the HCN(3-2) and HCO$^{+}$ spectra and confirmed it with the archived HCO$^{+}$(3-2) spectrum.   The most probable candidate for the absorber is a dense gas clump interacting with the radio jet, which was recently identified as the abrupt change in the hotspot motion by VLBI observations.  Alternatively, the absorber could be the outflowing gas expelled from the innermost CND by radiation pressure.  Time variability of the absorption feature would be detected if such an interaction is underway.  This will be verified with future monitoring observations. 

Using the observed velocity structure and archival {\it HST} data, we measured the dynamical mass of the SMBH.  The measured SMBH is $1\times10^{9}~M_{\sun}$, which agrees with the dynamical measurement using the warm H$_{2}$ gas but disagrees with the virial mass estimate using broad emission lines.  The discrepancies can be arisen by the complexity of BLR structure, which is possibly related to the violent time variation of AGN activity in NGC~1275.  Similar comparisons with other AGNs would be helpful to figure out the cause.  Increasing the sample of SMBH mass estimate by molecular gas dynamics with ALMA \citep[e.g.][]{Barth2016, Boizelle2019, Smith2019} is one of key approaches to shed light on this problem.

\bigskip 
We thank the referee for constructive comments.  This paper makes use of the following ALMA data: ADS/JAO.ALMA\#2017.0.01257.S and ADS/JAO.ALMA\#2013.1.01102.S. ALMA is a partnership of ESO (representing its member states), NSF (USA) and NINS (Japan), together with NRC (Canada), MOST and ASIAA (Taiwan), and KASI (Republic of Korea), in cooperation with the Republic of Chile. The Joint ALMA Observatory is operated by ESO, AUI/NRAO and NAOJ. HN is supported by JSPS KAKENHI Grant Number JP18K03709.  NK is supported by JSPS KAKENHI Grant Number JP16K17670.  YF is supported by JSPS KAKENHI Grant Number JP18K03647.  MK is supported by JSPS KAKENHI Grant Numbers JP18K03656 and JP18H03721.  KO acknowledges support from Shimadzu Science Foundation.

\renewcommand{\bibname}{}

\end{document}